\documentclass[graybox]{svmult}
\usepackage{amsmath}
\usepackage{mathptmx}       
\usepackage{helvet}         
\usepackage{courier}        
\usepackage{type1cm}        
\usepackage{makeidx}         
\usepackage{graphicx}        
\usepackage{multicol}        
\usepackage[bottom]{footmisc}
\usepackage{cite}

\newcommand{\cP}{{\cal P}}

\def\tr{\operatorname{tr\:}}

\makeindex             

\begin{document}

\title*{Anomalous Transport from Kubo Formulae}

\author{Karl Landsteiner, Eugenio Meg\'\i as and Francisco Pe\~na-Benitez}

\institute{Karl Landsteiner \at Instituto de F\'\i sica Te\'orica UAM-CISC, C/
Nicol\'as Cabrera 13-15, Univ. Aut\'onoma de Madrid, 28049 Madrid, Spain,
\email{karl.landsteiner@csic.es}
\and Eugenio Meg\'\i as \at Grup de F\'\i sica Te\`orica and IFAE, Departament
de F\'\i sica, Universitat Aut\`onoma de Barcelona, Bellaterra E-08193
Barcelona, Spain, 
\email{emegias@ifae.es}
\and Francisco Pe\~na-Benitez \at Instituto de F\'\i sica Te\'orica UAM-CISC, C/
Nicol\'as Cabrera 13-15 and \at Departamento de F\'isica Te\'orica, Univ.
Aut\'onoma de Madrid, 28049 Madrid, Spain,
\email{fran.penna@uam.es} }

\maketitle

\abstract*{Chiral anomalies have profound impact on the transport
  properties of relativistic fluids.  In four dimensions there are
  different types of anomalies, pure gauge and mixed gauge-gravitational
  anomalies. They give
  rise to two new non-dissipative transport coefficients, the chiral
  magnetic conductivity and the chiral vortical conductivity.  They
  can be calculated from the microscopic degrees of freedom with the
  help of Kubo formulae.  We review the calculation of the anomalous
  transport coefficients via Kubo formulae with a particular emphasis
  on the contribution of the mixed gauge-gravitational anomaly.}

\abstract{Chiral anomalies have profound impact on the transport
  properties of relativistic fluids.  In four dimensions there are
  different types of anomalies, pure gauge and mixed gauge-gravitational
  anomalies. They give
  rise to two new non-dissipative transport coefficients, the chiral
  magnetic conductivity and the chiral vortical conductivity.  They
  can be calculated from the microscopic degrees of freedom with the
  help of Kubo formulae.  We review the calculation of the anomalous
  transport coefficients via Kubo formulae with a particular emphasis
  on the contribution of the mixed gauge-gravitational anomaly.}

\section{Introduction}
\label{sec:intro}
Anomalies in relativistic field theories of chiral fermions belong to 
the most intriguing properties of quantum field theory. 
Comprehensive reviews on anomalies can be found in the
textbooks~\cite{Bertlmann:1996xk,Bastianelli:2006rx,Fujikawa:2004cx}.

Hydrodynamics is an ancient subject. Even in its
relativistic form it appeared that everything relevant to its
formulation could be found in \cite{Landau}. 
Apart from stability issues that were addressed in the 1960s and 1970s 
\cite{Muller:1967zza,Israel:1976tn,Israel2} leading to a second order formalism
there seemed little room for new discoveries. The last years witnessed however
an unexpected and profound development of the formulation of relativistic
hydrodynamics. 
The second order contributions have been put on a much more systematic basis
applying effective field theory
reasoning \cite{Baier:2007ix,fluidgravity}. The lessons learned from applying the AdS/CFT
correspondence \cite{Maldacena:1997re,Witten,GKP} to the plasma phase of strongly coupled
non-abelian gauge theories
\cite{Gubser:1996de, Witten:1998zw, Kovtun:2004de} played a major role (see
\cite{CasalderreySolana:2011us} for a recent review).

The presence of chiral anomalies in otherwise conserved currents has profound
implications for the formulation of relativistic hydrodynamics. 
The transport processes related to anomalies have surfaced several times and
independently \cite{Vilenkin:1979ui, Vilenkin2, Vilenkin3, Vilenkin4, Giovannini:1997eg,Alekseev:1998ds}. 
The axial current was the focus in \cite{Newman:2005as} and the first
application 
of the AdS/CFT correspondence
to anomalous hydrodynamics can be found already in~\cite{Newman:2005hd}. 
The full impact anomalies have on the formulation of relativistic hydrodynamics 
was however not fully appreciated until recently. 

The renewed interest in the formulation of relativistic hydrodynamics
has its origin mostly in the spectacular experimental evidence for
collective flow phenomena taking place in the physics of heavy ion
collisions at RHIC and LHC.  These experiments indicate the creation
of a deconfined quark gluon plasma in a strongly coupled regime.  In
the context of heavy ion collisions it was argued in
\cite{Kharzeev:2007jp,Fukushima:2008xe} that the excitation of
topologically non-trivial gluon field configurations in the early
non-equilibrium stages of a heavy ion collision might lead to an
imbalance in the number of left- and right-handed quarks. This
situation can be modeled by an axial chemical potential and it was
shown that an external magnetic field leads to an electric current
parallel to the magnetic field. This chiral magnetic effect leads then
to a charge separation perpendicular to the reaction plane in heavy
ion collisions. The introduction of an axial chemical potential also
allows to define a chiral magnetic conductivity which is simply the
factor of proportionality between the magnetic field and the induced
electric current.  This effect is a direct
consequence of the axial anomaly.

The application of the fluid/gravity correspondence to theories
including chiral anomalies lead to another surprise: it was found that
not only a magnetic field induces a current but that also a vortex in
the fluid leads to an induced current \cite{Erdmenger:2008rm,
  Banerjee:2008th}. This is the chiral vortical effect.  Again it is a
consequence of the presence of a chiral anomaly.  It was later
realized that the chiral magnetic and vortical conductivities are
almost completely fixed in the hydrodynamic framework by demanding the
existence of an entropy current with positive definite divergence
\cite{Son:2009tf}. That this criterion did not fix the anomalous
transport coefficients completely was noted in \cite{Neiman:2010zi}
and various terms depending on the temperature instead of the chemical
potentials were shown to be allowed as undetermined integration
constants. See also~\cite{Kalaydzhyan:2011vx,GKK} for a recent discussion
of these anomaly coefficients with applications to heavy ion physics.

In the meanwhile Kubo formulae for the chiral magnetic conductivity
\cite{Kharzeev:2009pj} and the chiral vortical conductivity \cite{Amado:2011zx}
had been developed. Up to this
point only pure gauge anomalies had been considered to be relevant since the
mixed gauge-gravitational anomaly in four dimensions is of higher order in
derivatives and was thought
not to be able to contribute to hydrodynamics at first order in derivatives.
Therefore it came as a surprise that in the application of the Kubo
formula for the 
chiral vortical conductivity to a system of free chiral fermions a purely
temperature dependent contribution was found. This contribution was consistent
with some the earlier found
integration constants and it was shown to arise if and only if the system of
chiral fermions features a mixed gauge-gravitational anomaly
\cite{Landsteiner:2011cp}. In fact these
contributions had been found already very early on in \cite{Vilenkin2,Vilenkin4}.
The connection to the presence of anomalies was however not made at that
time. The gravitational
anomaly contribution to the chiral vortical effect was also established in a
strongly coupled AdS/CFT approach and precisely the same result as at weak
coupling was found \cite{Landsteiner:2011iq}. 

The argument based on a positive definite divergence of the entropy
current allows to fix the contributions form pure gauge anomalies uniquely and
provides therefore a non-renormalization theorem. 
No such result is known thus far for the
contributions of the gauge-gravitational anomaly.~\footnote{See however the very
recent attempts to establish non-renormalization theorems in
\cite{Golkar:2012kb} and \cite{Jensen:2012kj}.} 

A gas of weakly coupled Weyl fermions in arbitrary dimensions has been studied
in \cite{Loganayagam:2012pz} and confirmed that the anomalous conductivities can
be obtained
directly from the anomaly polynomial under substitution of the field strength
with the chemical potential and the first Pontryagin density by the negative of
the temperature squared \cite{Loganayagam:2011mu} . Recently the anomalous
conductivities have
also been obtained in effective action approaches
\cite{Jensen:2012jy,Banerjee:2012iz}. The contribution of the mixed
gauge-gravitational
anomaly appear on all these approaches as undetermined integrations constants.

We will review here what can be learned from the calculation of the anomalous
conductivities via Kubo formulae. The advantage of the usage of Kubo formulae
is that they capture all contributions stemming either from pure gauge or from
mixed gauge-gravitational anomalies. The disadvantage is that the calculations
can
be performed only with a particular model and only in a weak or in the
gravity dual of the strong coupling regime. 
Along the way we will explain our 
point of view on some subtle issues concerning the definition of currents and
of chemical potentials when anomalies are present. These subtleties lead indeed
to some ambiguous
results \cite{Yee:2009vw} and \cite{Rebhan:2009vc}. A first step to
clarify these issues
was done in \cite{Gynther:2010ed} and a more general exposition of the
relevant issues has appeared in \cite{Landsteiner:2011tf}. 

The review is organized as follows. In section two we will briefly
summarize the relevant issues concerning anomalies. We recall how
vector like symmetries can always be restored by adding suitable
finite counterterms to the effective action~\cite{Bardeen:1969md}. A
related but different issue is the fact that currents can be defined
either as consistent or as covariant currents. The hydrodynamic
constitutive relations depend on what definition of current is used.
We discuss subtleties in the definition of the chemical potential in
the presence of an anomaly and define our preferred scheme. We discuss
the hydrodynamic constitutive relations and derive the Kubo formulae
that allow the calculation of the anomalous transport coefficients
from two point correlators of an underlying quantum field theory.

In section three we apply the Kubo formulae to a theory of free Weyl fermions
and show that two different contributions arise. They are clearly identifiable 
as being related to the presence of pure gauge and mixed gauge-gravitational
anomalies.

In section four we define a holographic model that implements the mixed
gauge-gravitational anomaly via a mixed gauge-gravitational Chern-Simons term.
We calculate the same Kubo formulae as at weak
coupling, obtaining the same values for chiral axi-magnetic and chiral vortical
conductivities as in the weak coupling model.

We conclude this review with some discussions and outlook to further
developments.

\section{Anomalies and Hydrodynamics}
\label{sec:2}

In this section we will review  briefly anomalies. 
We compare the consistent with the covariant form of the anomaly and
we introduce the Bardeen
counterterm that allows to restore current conservation for vector like
symmetries. Then we turn to the question of what we mean when we talk about the
chemical potential.
Two ways of introducing chemical potential, either through a deformation of the
Hamiltonian or by demanding twisted boundary conditions along the thermal circle
are shown
to be in-equivalent in presence of an anomaly. Equivalence can still be achieved
by introduction of a spurious axion field. We explain the implications for
holography.
The constitutive relations for anomalous currents are introduced in Landau and
Laboratory frame. We discuss how they differ if we use the consistent instead
of the covariant currents and derive the Kubo formulae for the anomalous
conductivities.

\subsection{Anomalies}
\label{subsec:2}
Anomalies arise by integrating over
chiral fermions in the path integral. They signal a fundamental 
incompatibility between the symmetries present in the classical 
theory and the quantum theory.

Unless otherwise stated we will always
think of the symmetries as global symmetries. 
But we still will introduce gauge fields. These gauge
fields serve as classical sources coupled to the currents. 
As a side effect their presence promotes the global symmetry to a local gauge
symmetry. It is still justified to think of it as a global symmetry as long as
we do not introduce a kinetic Maxwell or Yang-Mills term in the action. 

In a theory with chiral fermions we define an effective action depending on
these gauge fields by the path integral
\begin{equation}
 \label{eq:Weff}
 e^{i W_{eff}[A_\mu]/\hbar } := \int {\cal D}\Psi {\cal D}\bar \Psi e^{ i
S[\psi,
A_\mu]/\hbar }\,.
\end{equation}
The vector field $A_\mu(x)$ couples to a classically conserved current $J^\mu =
\bar\Psi
\gamma^\mu Q \Psi$. The charge operator $Q$ can be the generator of a Lie group
combined with
chiral projectors $\mathcal P_{\pm} = \frac{ 1}{ 2} (1\pm\gamma_5)$. General
combinations are
allowed although in the following we will mostly concentrate on a simple chiral
$U(1)$ symmetry 
for which we can take $Q=\mathcal P_+$. The fermions are minimally
coupled to the gauge field and the classical action has an underling gauge
symmetry 
\begin{equation}
 \label{eq:symclassical}
 \delta \Psi = -i \lambda(x) Q \Psi    ~~~~~~~~~~~~~~,~~~~~~~~~~~~ \delta
A_\mu(x) = D_\mu \lambda(x) \,,
\end{equation}
with $D_\mu$ denoting the gauge covariant derivative. 
Assuming that the theory has a classical limit the effective action in terms of
the gauge fields allows for an expansion in $\hbar$
\begin{equation}
 \label{eq:expansionWeff}
 W_{eff} = W_0 + \hbar W_1 + \hbar^2 W_2 + \dots
\end{equation}
We find it convenient to use the language of BRST symmetry by promoting the gauge
parameter to a ghost field $c(x)$.~\footnote{A recent comprehensive review on BRST
symmetry 
is \cite{Dragon:2012au}.}
The BRST symmetry is generated by
\begin{equation}
 \label{eq:BRST}
  s A_\mu = D_\mu c   ~~~~~~, ~~~~~ sc = -i c^2\,.
\end{equation}
It is nilpotent $s^2=0$. 
The statement that the theory has an anomaly can now be neatly formalized. Since
on gauge fields the BRST symmetry acts just as the gauge symmetry,  gauge
invariance
translates into BRST invariance. An anomaly is present if
\begin{equation}
 \label{eq:anomaly}
 s W_{eff} = {\cal A}  ~~~~~~~~ \mathrm{and} ~~~~ {\cal A} \neq sB \,.
\end{equation}
Because of the nilpotency of the BRST operator the anomaly has to fulfill the
Wess-Zumino consistency condition 
\begin{equation}
 s{\cal A}=0\,.
\end{equation}
As indicated in (\ref{eq:anomaly}) this has a possible trivial
solution if there exists a local functional $B[A_\mu]$ such that $sB =
{\cal A}$. An anomaly is present if no such $B$ exists.  The anomaly
is a quantum effect. If it is of order $\hbar^n$ and if a suitable
local functional $B$ exists we could simply redefine the effective
action as $\tilde{W}_{eff} = W_{eff} - B$ and the new effective action
would be BRST and therefore gauge invariant. The form and even the
necessity to introduce a functional $B$ might depend on the particular
regularization scheme chosen. As we will explain in the case of an
axial and vector symmetry a suitable $B$ can be found that always
allows to restore the vectorlike symmetry, this is the so-called
Bardeen counterterm \cite{Bardeen:1969md}. The necessity to introduce
the Bardeen counterterm relies however on the regularization scheme
chosen. In schemes that automatically preserve vectorlike symmetries,
such as dimensional regularization, the vector symmetries are
automatically preserved and no counterterm has to be
added. Furthermore the Adler-Bardeen theorem guarantees that chiral
anomalies appear only at order $\hbar$. Their presence can therefore
by detected in one loop diagrams such as the triangle diagram of three
currents.

We have introduced the gauge fields as sources for the currents
\begin{equation}
 \label{eq:sources}
 \frac{\delta}{\delta A_\mu(x)} W_{eff}[A] = \langle J^\mu \rangle\,.
\end{equation}
For chiral fermions transforming under a general Lie group generated by $T^a$
the chiral anomaly takes the form \cite{Bertlmann:1996xk}
\begin{eqnarray}
\nonumber s W_{eff}[A] &=& -\int d^4x c^a (D_\mu J^\mu)^a \\
\label{eq:anomalyform} &=&  -\frac{\eta}{24\pi^2}\int d^4x\, c^a
\epsilon^{\mu\nu\rho\sigma} \mathrm{tr} \left[ T^a \partial_\mu
\left( A_\nu \partial_\rho A_\sigma + \frac{ 1}{ 2} A_\nu A_\rho A_\sigma
\right)
\right]  \,.
\end{eqnarray}
Where $\eta=+1$ for left-handed fermions and $\eta=-1$ for
right-handed fermions.  Differentiating with respect to the ghost
field (the gauge parameter) we can derive a local form. To simplify
the formulas we specialize this to the case of a single chiral $U(1)$
symmetry taking $T^a=1$
\begin{equation}
 \label{eq:localanomaly}
 \partial_\mu J^\mu = \frac{\eta}{96\pi^2} \epsilon^{\mu\nu\rho\sigma}
F_{\mu\nu}F_{\rho\sigma}\,.
\end{equation}
This is to be understood as an operator equation. Sandwiching it between the
vacuum state $\left|0\right>$ and further
differentiating with respect to the gauge fields we can generate the famous
triangle
form of the anomaly
\begin{equation}
 \label{eq:triangleanomaly}
 \langle \partial_\mu J^\mu(x) J^\sigma(y) J^\kappa(z) \rangle =
\frac{1}{12\pi^2}\epsilon^{\mu\sigma\rho\kappa} \partial^x_\mu \delta(x-y)
\partial^x_\rho \delta(x-z) \,.
\end{equation}
The one point function of the divergence of the current is non-conserved only in
the background of parallel electric and magnetic fields whereas the
non-conservation of the current
as an operator becomes apparent in the triangle diagram even in vacuum. 

By construction this form of the anomaly fulfills the Wess-Zumino
consistency condition and is therefore called the {\em consistent
  anomaly}. In analogy we call the current defined by
(\ref{eq:sources}) the {\em consistent current}.

For a $U(1)$ symmetry the functional differentiation with respect to the gauge
field  and the BRST operator $s$ commute, 
\begin{equation}
 \label{eq:scommute}
\left[s, \frac{\delta}{\delta A_\mu(x)} \right] =0\,.
\end{equation}
An immediate consequence is that the consistent current is not BRST invariant
but rather obeys
\begin{equation}
 \label{eq:constentcurrentstrafo}
s J^\mu = \frac{1}{24\pi^2} \epsilon^{\mu\nu\rho\lambda} \partial_\nu c
F_{\rho\lambda} = - \frac{1}{24\pi^2} s K^\mu \,,
\end{equation}
where we introduced the Chern-Simons current $K^\mu =
\epsilon^{\mu\nu\rho\lambda} A_\nu F_{\rho\lambda}$ with $\partial_\mu
K^\mu = \frac{ 1}{ 2} \epsilon^{\mu\nu\rho\lambda} F_{\mu\nu}
F_{\rho\lambda} $. 

With the help of the Chern-Simons current it is
possible to define the so-called {\em covariant current} (in the case
of a $U(1)$ symmetry rather the invariant current)
\begin{equation}
 \label{eq:covcurrent}
\tilde J^\mu = J^\mu + \frac{1}{24\pi^2} K^\mu\,.
\end{equation}
fulfilling 
\begin{equation}
 s\tilde J^\mu =0\,.
\end{equation}

The divergence of the covariant current defines the {\em covariant anomaly}
\begin{equation}
 \label{eq:covanomaly}
 \partial_\mu \tilde{J}^\mu = \frac{1}{32\pi^2} \epsilon^{\mu\nu\rho\sigma}
F_{\mu\nu}F_{\rho\sigma}\,.
\end{equation}
Notice that the Chern-Simons current cannot be obtained as the variation with
respect to the gauge field of any funtional. It is
therefore not
possible to define an effective action whose derivation with respect to the
gauge field gives the covariant current.

Let us suppose now that we have one left-handed and one right-handed
fermion with the corresponding left- and right-handed
anomalies. Instead of the left-right basis it is more convenient to
introduce a vector-axial basis by defining the vectorlike current
$J^\mu_V = J^\mu_L + J^\mu_R$ and the axial current $J_A^\mu =
J^\mu_L-J^\mu_R$.  Let $V_\mu(x)$ be the gauge field that couples to
the vectorlike current and $A_\mu(x)$ be the gauge field coupling to
the axial current. The (consistent) anomalies for the vector and axial
current turn out to be
\begin{eqnarray}
 \label{eq:vectoranomaly}
 \partial_\mu J^\mu_V &=& \frac{1}{24\pi^2} \epsilon^{\mu\nu\rho\lambda}
F_{\mu\nu}^V F_{\rho\lambda}^A \,,\\
\label{eq:axialanomaly}
 \partial_\mu J^\mu_A &=& \frac{1}{48\pi^2} \epsilon^{\mu\nu\rho\lambda}
(F_{\mu\nu}^V F_{\rho\lambda}^V + F_{\mu\nu}^A F_{\rho\lambda}^A) \,.
\end{eqnarray}
As long as the vectorlike current corresponds to a global symmetry
nothing has gone wrong so far. If we want to identify the vectorlike
current with the electricmagnetic current in nature we need to couple
it to a dynamical photon gauge field and now the non-conservation of
the vector current is worrysome to say the least. The problem arises
because implicitly we presumed a regularization scheme that treats
left-handed and right-handed fermions on the same footing.  As pointed
out first by Bardeen this flaw can be repaired by adding a finite
counterterm (of order $\hbar$) to the effective action. This is the
so-called Bardeen counterterm and has the form
\begin{equation}
 \label{eq:Bardeenct}
 B_{ct} = -\frac{1}{12\pi^2} \int d^4x \,\epsilon^{\mu\nu\rho\lambda} V_\mu
A_\nu
F_{\rho\lambda}^V \,.
\end{equation}
Adding this counterterm to the effective action gives additional contributions
of Chern-Simons form to the consistent vector and axial currents. With the
particular
coefficient chosen it turns out that the anomaly in the vector current is
canceled whereas the axial current picks up an additional contribution such that
after
adding the Bardeen counterterm the anomalies are
\begin{eqnarray}
 \label{eq:vectorconserved}
 \partial_\mu J^\mu_V &=&0 \,,\\
\label{eq:bardeenaxialanomaly}
 \partial_\mu J^\mu_A &=& \frac{1}{48\pi^2} \epsilon^{\mu\nu\rho\lambda} (3
F_{\mu\nu}^V F_{\rho\lambda}^V + F_{\mu\nu}^A F_{\rho\lambda}^A) \,.
\end{eqnarray}
This definition of currents is mandatory if we want to identify the vector
current with the usual electromagnetic current in nature! It is furthermore
worth
to point out that both currents are now invariant under the vectorlike $U(1)$ 
symmetry. The currents are not invariant under axial transformation, but these
are anomalous anyway.

Generalizations of the covariant anomaly and the Bardeen counterterm to the
non-abelian case can be found e.g. in \cite{Bertlmann:1996xk}.

There is one more anomaly that will play a major role in this review, the mixed
gauge-gravitational anomaly \cite{Delbourgo:1972xb, Delbourgo2, Eguchi}.~\footnote{In $D=4k+2$
dimensions 
also purely gravitational anomalies can appear \cite{AlvarezGaume:1983ig}.} 
So far we have considered only spin one currents
and have coupled them to gauge fields. Now we also want to introduce the
energy-momentum tensor through its coupling to a fiducial background metric
$g_{\mu\nu}$. 
Just as the gauge fields, the metric serves primarily as the source for
insertions of the energy momentum tensor in correlation functions. Just as in
the case
of vector and axial currents, the mixed gauge-gravitational anomaly is the
statement that it is impossible in the quantum theory to preserve at the same
time the
vanishing of the divergence of the energy-momentum tensor and of chiral (or
axial) $U(1)$ currents. It is however possible to add Bardeen counterterms to
shift
the anomaly always in the sector of the spin one currents and preserve
translational (or diffeomorphism) symmetry. If we have a set of left-handed and
right-handed
chiral fermions transforming under a Lie Group generated by $(T_a)_L$ and
$(T_a)_R$ in the background of arbitrary gauge fields and metric, the anomaly
is conveniently expressed through the non-conservation of the {\em covariant} 
current as\footnote{We thank A. Grushin for pointing out a crucial mistake in a previous arXiv version.}
\begin{eqnarray}
 D_\mu T^{\mu\nu} &=&  F^{\nu\mu}_a J^a_{\mu} + 
 \frac{b_a}{384\pi^2} D_\mu \left(\epsilon^{\sigma\kappa\rho\lambda} F^a_{\kappa\sigma} R^{\nu\mu}\,_{\rho\lambda} \right)\,,\\
 \label{eq:gravanomaly}
 (D_\mu J^\mu)_a &=& \frac{d_{abc}}{32\pi^2} \epsilon^{\mu\nu\rho\lambda}
F^b_{\mu\nu} F^c_{\rho\lambda} + \frac{b_a}{768\pi^2}
\epsilon^{\mu\nu\rho\lambda} R^\alpha\,_{\beta \mu\nu}
R^\beta\,_{\alpha \rho\lambda}\,.
\end{eqnarray}
The purely group theoretic factors are
\begin{eqnarray}
 d_{abc} &=& \frac{ 1}{ 2} \mathrm{tr} ( T_a \{ T_b , T_c\} )_L - \frac{ 1}{ 2}
\mathrm{tr} ( T_a \{ T_b , T_c\} )_R \,, \label{eq:chiralcoeff} \\ 
 b_a &=& \mathrm{tr}(T_a)_L - \mathrm{tr}(T_a)_R\,. \label{eq:gravcoeff}
\end{eqnarray}
Chiral anomalies are completely absent if and only if $d_{abc}=0$ and $b_a$=0.


\subsection{Chemical Potentials for anomalous symmetries}
\label{subsec:chempot}
Thermodynamics of systems with conserved charges can be described in a grand
canonical ensemble where a Lagrange multiplier $\mu$ ensures that the 
partition function fulfills
\begin{equation}
 \label{eq:chempotdef}
 T\frac{\partial \log(Z)}{\partial \mu} =  \langle Q \rangle\,.
\end{equation}
The textbook approach is to consider a deformation of the Hamiltonian
\begin{equation}\label{eq:deformedH}
 H \rightarrow H - \mu Q \,,
\end{equation}
where $Q$ is the charge in question. We can think of this as arising
from the coupling of the (fiducial) gauge field $A_\mu$ to the current
$J^\mu$  and giving a vacuum
expectation value to $A_0 = \mu$. Since the fiducial gauge field leads
to local gauge invariance we can remove the $\mu Q$
coupling in the Hamiltonian by the gauge transformation $A_0
\rightarrow A_0 + \partial_0 \chi$ with $\chi = - \mu t$. 

\begin{figure}

\scalebox{0.5}{\setlength{\unitlength}{4144sp}%
\begingroup\makeatletter\ifx\SetFigFont\undefined%
\gdef\SetFigFont#1#2#3#4#5{%
  \reset@font\fontsize{#1}{#2pt}%
  \fontfamily{#3}\fontseries{#4}\fontshape{#5}%
  \selectfont}%
\fi\endgroup%
\begin{picture}(8857,3265)(2416,-4607)
\thicklines
{\color[rgb]{0,0,0}\put(2701,-1861){\line( 1, 0){8550}}
\put(11251,-1861){\line( 0,-1){225}}
\put(11251,-2086){\line(-1, 0){8550}}
\put(2701,-2086){\line( 0,-1){2025}}
}%
\put(11026,-1726){\makebox(0,0)[lb]{\smash{{\SetFigFont{20}{24.0}{\familydefault}{\mddefault}{\updefault}{\color[rgb]{0,0,0}$t_f$}%
}}}}
\put(2476,-1681){\makebox(0,0)[lb]{\smash{{\SetFigFont{20}{24.0}{\familydefault}{\mddefault}{\updefault}{\color[rgb]{0,0,0}$t_i$}%
}}}}
\put(2431,-4471){\makebox(0,0)[lb]{\smash{{\SetFigFont{20}{24.0}{\familydefault}{\mddefault}{\updefault}{\color[rgb]{0,0,0}$t_i-i\beta$}%
}}}}
\end{picture}%
 }

\caption{At finite temperature field theories are defined on the
Keldysh-Schwinger contour in the complexified time plane. The initial stated
at $t_i$ is specified through the boundary conditions on the fields. The
endpoint of the contour is at $t_i-i\beta$ where $\beta=1/T$.}
\label{fig:KeldyshSchwinger}       
\end{figure}
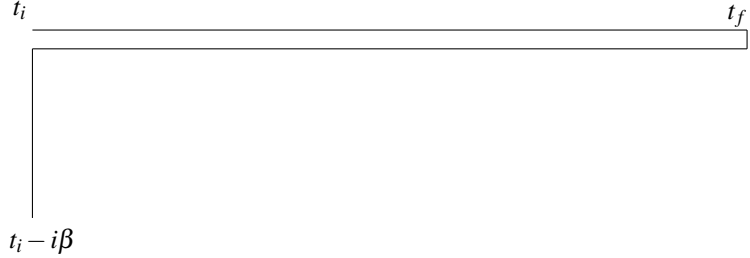

In the context of finite temperature field theory such a gauge transformation is
however not really 
allowed. One needs to define the field theory on the Keldysh-Schwinger contour
in the complexified time
plane as shown in figure (\ref{fig:KeldyshSchwinger}). Fields are taken to be
periodic or anti-periodic 
along the imaginary time direction $t= -i \tau$ with period $\beta=1/T$ where
$T$ is the temperature
\begin{equation}
\label{eq:bcs1}
 \Psi(t_i-i\beta) =  \pm \Psi(t_i)\,,
\end{equation}
with the plus sign for bosons and the minus sign for fermions.
The gauge transformation that removes the constant zero component of the gauge
field is not periodic
along the contour and therefore changes the boundary conditions on the fields.
After the gauge transformation with $\chi=-\mu t$ the fields obey
the  boundary conditions
\begin{equation}
\label{eq:bcs2}
 \Psi(t_i-i\beta) =  \pm e^{q \mu \beta} \Psi(t_i)\,.
\end{equation}

Demanding these ``twisted'' boundary conditions is of course completely
equivalent to having $A_0=\mu$. The gauge invariant
statement is that a charged field parallel transported around the
Keldysh-Schwinger contour picks up a factor of
$\exp(q\mu\beta)$. 
As long as we have honest non-anomalous symmetries
under consideration we have therefore two (gauge)-equivalent
formalisms of how to introduce the chemical potential summarized in
table~\ref{tab:formalisms} \cite{Evans:1995yz}.

\begin{table}
\caption{Two formalisms for the chemical potential}
\label{tab:formalisms}       
%
%
\begin{tabular}{p{2cm}p{2.4cm}p{4cm}}
\hline\noalign{\smallskip}
Formalism  & Hamiltonian & Boundary condition \\
\noalign{\smallskip}\svhline\noalign{\smallskip}
(A) &  $H-\mu Q$  & $\Psi(t_i-i\beta) = \pm  \Psi(t_i)$ \\
(B) & $H$ & $\Psi(t_i-i\beta) = \pm e^{q \beta \mu} \Psi(t_i)$\\
\noalign{\smallskip}\hline\noalign{\smallskip}
\end{tabular}
\end{table}

One convenient point of view on formalism (B) is the following. In a
real time Keldysh-Schwinger setup we demand some initial conditions at
initial (real) time $t=t_i$. These initial conditions are given by the
boundary conditions in (B). From then on we do the (real) time
development with the microscopic Hamiltonian $H$. In principle there
is no need for the Hamiltonian $H$ to preserve the symmetry present
at times $t<t_i$. This seems an
especially suited approach to situations where the charge in question
is not conserved by the real time dynamics. In the case of an
anomalous symmetry we can start at $t=t_i$ with a state of certain
charge. As long as we have only external gauge fields present the
one-point function of the divergence of the current vanishes and
the charge is conserved. This is not true on the full theory
since even in vacuum the three-point correlators are sensitive to
the anomaly. For the formulation of hydrodynamics in external fields
the condition that the one-point functions of the currents are conserved
as long as there are no parallel electric and magnetic external fields (or
a metric that has non-vanishing Pontryagin density) is sufficient.~\footnote{If
dynamical gauge fields are present, such as the gluon fields in QCD
even the one point function of the charge does decay over (real) time due to
non-perturbative processes (instantons) or at finite temperature due to
thermal sphaleron processes \cite{Moore:2010jd}. Even in this case
in the limit of large number of colors these processes are suppressed and
can e.g. not be seen in holographic models in the supergravity approximation.}

Let us assume now that $Q$ is an anomalous charge, i.e. its 
associated current suffers from chiral anomalies.
We first consider formalism (B) and ask what happens if we do now the gauge
transformation that would bring us to formalism (A). Since the
symmetry is anomalous the action transforms as
\begin{equation}
S[A+\partial\chi] = S[A] + \int d^4x\, \chi
\epsilon^{\mu\nu\rho\lambda}\left(C_1 F_{\mu\nu}F_{\rho\lambda} + C_2
R^\alpha\,_{\beta\mu\nu} R^\beta\,_{\alpha\rho\lambda}\right) \,,
\end{equation}
with the anomaly coefficients $C_1$ and $C_2$ depending on the chiral
fermion content. It follows that formalisms (A) and (B) are physically
inequivalent now, because of the anomaly. However, we would like to
still come as close as possible to the formalism of (A) but in a form
that is physically equivalent to the formalism (B). To achieve this we
proceed by introducing a non-dynamical axion field $\Theta(x)$ and the
vertex
\begin{equation}\label{eq:Stheta}
S_\Theta[A,\Theta] = \int d^4x \, \Theta\, \epsilon^{\mu\nu\rho\lambda}\left(C_1
F_{\mu\nu}F_{\rho\lambda} + C_2 R^\alpha\,_{\beta\mu\nu}
R^\beta\,_{\alpha\rho\lambda}\right) \,.
\end{equation}
If we demand now that the ``axion'' transforms as $\Theta \rightarrow \Theta -
\chi$ under gauge transformations
we see that the action
\begin{equation}
S_{tot}[A,\Theta] = S[A] + S_\Theta[A,\Theta] 
\end{equation}
is gauge invariant. Note that this does not mean that the theory is
not anomalous now. We introduce it solely for the purpose to make
clear how the action has to be modified such that two field
configurations related by a gauge transformation are physically
equivalent. It is better to consider $\Theta$ as {\em coupling} and
not a field, i.e. we consider it a spurion field.  The gauge field
configuration that corresponds to formalism (B) is simply $A_0=0$. A
gauge transformation with $\chi=\mu t$ on the gauge invariant action
$S_{tot}$ makes clear that a physically equivalent theory is obtained
by chosing the field configuration $A_0=\mu$ and the time dependent
coupling $\Theta = -\mu t$. If we define the current through the
variation of the action with respect to the gauge field we get an
additional contribution from $S_\Theta$,
\begin{equation}
J_\Theta^\mu = 4 C_1 \epsilon^{\mu\nu\rho\lambda} \partial_\nu \Theta
F_{\rho\lambda}\,, 
\end{equation}
and evaluating this for $\Theta= -\mu t$ we get the spatial current
\begin{equation}\label{eq:Jtheta}
J_\Theta^m = 8 C_1 \mu B^m\,.
\end{equation}
We do not consider this to be the chiral magnetic effect! This is only the
contribution to the current that comes from the new coupling that we
are forced to introduce by going to formalism (A) from (B) in a
(gauge)-equivalent way. As we will see in the following
chapters the chiral magnetic and vortical effect are
on the contrary non-trivial results of dynamical one-loop calculations. 

What is the Hamiltonian now based on
the modified formalism (A)? We have to take of course the new coupling
generated by the non-zero $\Theta$. The Hamiltonian now is therefore
\begin{equation}\label{eq:Hmodified}
H - \mu \left( Q + 4 C_1 \int d^3x \epsilon^{0ijk} A_i \partial_j A_k\right)\,, 
\end{equation}
where for simplicity we have ignored the contributions from the metric terms.

For explicit computations in sections~\ref{sec:3} and \ref{sec:4} we will
introduce the chemical potential through the
formalism (B) by demanding twisted boundary conditions. It seems the
most natural choice since the dynamics is described by the microscopic
Hamiltonian $H$. The modified (A) based on the Hamiltonian
(\ref{eq:Hmodified}) is however not without merits. It is convenient
in holography where it allows vanishing temporal gauge field on the
black hole horizon and therefore a non-singular Euclidean black hole
geometry.~\footnote{It is possible to define a generalized formalism
  to make any choice for the gauge field $A_0=\nu$, so that one
  recovers formalism (A) when $\nu=\mu$ and formalism (B) when $\nu=0$
  as particular cases (see~\cite{Megias:2012Padua} for details).}

\subsubsection{Hydrodynamics and Kubo formulae}
\label{subsec:hydro}
The modern understanding of hydrodynamics is as an effective field theory. The
equations of motion are the 
(anomalous) conservation laws of the energy-momentum tensor and spin one
currents. These are supplemented by
expression for the energy-momentum tensor and the current which are organized in
a derivative expansion, the 
so-called constitutive relations. Symmetries constrain the possible terms. In
the presence of chiral anomalies the 
constitutive relations for the energy-momentum tensor and the currents in the
Landau frame are
\begin{eqnarray}\label{eq:Landauframe1}
 T^{\mu\nu} &=& \epsilon u^\mu u^\nu + p P^{\mu\nu} -\eta
P^{\mu\alpha}P^{\nu\beta} \sigma_{\alpha \beta} -\zeta P^{\mu\nu}
\partial_\alpha u^\alpha \,,\\
\label{eq:Landauframe2}
\tilde J^\mu_a &=& \rho_a u^\mu + \Sigma_{ab} \left( E^\mu_b - T
P^{\mu\alpha}D_\alpha\frac{\mu_a}{T}\right) +  \xi^B_{ab} B^\mu_a  + \xi^V_a
\omega^\mu \,.
\end{eqnarray}
It is important to specify that these are the constitutive relations
for the {\em covariant} currents!  Here $\epsilon$ is the energy
density, $p$ the pressure density, $u^\mu$ the local fluid
velocity. $P^{\mu\nu}= g^{\mu\nu} + u^\mu u^\nu$ is the transverse
projector to the fluid velocity. $\sigma_{\mu\nu}$~is the symmetric
traceless shear tensor.  The non-anomalous transport coefficients are
the shear viscosity $\eta$, the bulk viscosity $\zeta$ and the
electric conductivities $\Sigma_{ab}$. External electric and magnetic
fields are covariantized via $E^\mu_a = F_a^{\mu\nu}u_\nu$ and
$B^\mu_a = \frac 1 2 \epsilon^{\mu\nu\rho\lambda} u_\nu
F_{a,\rho\lambda}$. The vorticity of the fluid is 
$\omega^\nu = \epsilon^{\mu\nu\rho\lambda} u_\nu \partial_\lambda u_\rho$.

The anomalous transport coefficients are the chiral magnetic
conductivities $\xi^{B}_{ab}$ and the chiral vortical conductivities
$\xi^V_a$.  At first order in derivatives the notion of fluid velocity
is ambiguous and needs to be fixed by prescribing a choice of
frame. We remark that the constitutive relations
(\ref{eq:Landauframe1}) and (\ref{eq:Landauframe2}) are valid in the
Landau frame where $T^{\mu\nu}u_\nu = \epsilon u^\mu$.

To compute the Kubo formulae for the anomalous transport coefficients it turns
out that the Landau frame is not the most convenient one.
It fixes the definition of the fluid velocity through energy transport.
Transport phenomena related to the generation of an energy current
are therefore not directly visible, rather they are absorbed in the definition
of the fluid velocity. It is therefore more convenient to 
go to another frame in which we demand that the definition of the fluid velocity
is not influenced when switching on an external magnetic
field or having  a vortex in the fluid. In such a frame the constitutive
relations take the form
\begin{eqnarray}\label{eq:alternativeframe1}
\!\!\!\!\!\! T^{\mu\nu} &=& (\epsilon+p)u^\mu u^\nu + p g^{\mu\nu} - \eta
P^{\mu\alpha}P^{\nu\beta}\sigma_{\alpha\beta} - \zeta P^{\mu\nu} \partial_\alpha
u^\alpha   + Q^\mu u^\nu + Q^\nu u^\mu,\\
 \label{eq:alternativeframe2}
\!\!\!\!\!\! Q^\mu &=& \sigma^\epsilon_B B^\mu + \sigma^\epsilon_V
\omega^\mu\,,\\
\label{eq:alternativeframe3}
\!\!\!\!\!\! \tilde J^\mu &=& \rho u^\mu + \Sigma \left( E^\mu - T
P^{\mu\alpha}D_\alpha\left(\frac{\mu}{T}\right)\right) + \sigma_B B^\mu +
\sigma_V \omega^\mu\,.
\end{eqnarray}
In order to avoid unnecessary clutter in the equations we have specialized now
to a single $U(1)$ charge. Notice that now there is a sort of ``heat'' current
present in the constitutive relation for the energy-momentum tensor.

The derivation of Kubo formulae is better based on the usage of the
{\em consistent} currents. Since the covariant and consistent currents
are related by adding suitable Chern-Simons currents the constitutive
relations for the consistent current receives additional contribution
from the Chern-Simons current
\begin{equation}\label{eq:consistentconstitutive}
 J^\mu = \tilde J^\mu - \frac{1}{24\pi^2} K^\mu\,.
\end{equation}
If we were to introduce the chemical potential according to formalism
$(A)$ via a background for the temporal gauge field we would get an
additional contribution to the consistent current from the
Chern-Simons current. In this case it is better to go to the modified
formalism $(A')$ that also introduces a spurious axion field and
another contribution to the current $J_\Theta$ (\ref{eq:Jtheta}) has
to be added
\begin{equation}\label{eq:thetaconstitutive}
 J^\mu = \tilde J^\mu - \frac{1}{24\pi^2} K^\mu + J^\mu_\Theta\,.
\end{equation}

For the derivation of the Kubo formulae it is therefore more convenient to work
with formalism $(B)$ in which $A_0=0$ and the chemical potential
is introduced via the boundary conditions (\ref{eq:bcs2}). Otherwise there arise
additional contributions to the two point functions. We will briefly discuss
them in the next subsection.

From the microscopic view the constitutive relations should be interpreted as
the one-point functions of the operators $T^{\mu\nu}$ and $J^\mu$ in a
near equilibrium situation, i.e. gradients in the fluid velocity, the
temperature or the chemical potentials are assumed to be small. From this
point of view Kubo formulae can be derived. In the microscopic theory the
one-point function of an operator near equilibrium is given by linear
response theory whose basic ingredient are the retarded two-point functions. If
we consider a situation with only an external electric field in 
$z$-direction and
all other sources switched off, i.e. the fluid being at rest $u^\mu=(1,0,0,0)$
and no gradients in temperature or chemical potentials the constitutive
relations are simplified to 
\begin{equation}
 J^z = \Sigma E^z \,.
\end{equation}
The electric field is $E^z=i\omega A^z$ in terms of the vector potential and
using linear response theory the induced current is given 
through the retarded two-point function by
\begin{equation}
 J^z = \left\langle J^z J^z \right\rangle A^z\,.
\end{equation}
Equating the two expressions for the current we find the Kubo formula for the
electric conductivity
\begin{equation}\label{eq:sigmaelectric}
 \Sigma = \lim_{\omega\rightarrow 0} \frac{-i}{\omega} \left\langle J^z J^z
\right\rangle \,.
\end{equation}
This has to be evaluated at zero momentum. The limit in the frequency follows
because the constitutive relation are supposed to be valid
only to lowest order in the derivative expansion, therefore one needs to isolate
the first non-trivial term.

Now we want to find some simple special cases that allow the derivation of
Kubo
formulae for the anomalous conductivities. A very convenient choice is
to go to the restframe $u^\mu=(1,0,0,0)$, switch on a vector potential in the
$y$-direction that depends only on the $z$ direction and at the same time
a metric deformation $g_{\mu\nu} = \eta_{\mu\nu} + h_{\mu\nu}$ with the only
non-vanishing component $h_{0y}$ depending on $z$ only.
To linear order in the background fields the non-vanishing components of the
energy-momentum tensor and the current are
\begin{eqnarray}
 T^{0x} &=& -\sigma^\epsilon_B \partial_z A_y   -\sigma^\epsilon_V \partial_z
h_{0y}\,,\\
 J^x &=& -\sigma_B \partial_z A_y   -\sigma_V \partial_z h_{0y}\,,
\end{eqnarray}
since in the formalism $(B)$ neither the Chern-Simons term nor the $\Theta$
coupling contribute.
Going to momentum space and differentiating with respect to the sources $A_y$
and $h_{0y}$ we find therefore the Kubo formulae
\cite{Fukushima:2008xe,Amado:2011zx}

\begin{equation}\label{Kformulae}
\boxed{ \begin{array}{lcl}
&&\\
  \sigma_B = \lim_{k_z\rightarrow 0}  \frac i k_z \langle J^x J^y \rangle ~~~~
&& ~~~~\sigma_V = \lim_{k_z\rightarrow 0}  \frac i k_z \langle J^x T^{0y}
\rangle \\ & \\
  \sigma^\epsilon_B = \lim_{k_z\rightarrow 0}  \frac i k_z \langle T^{0x} J^y
\rangle ~~~~
&& ~~~~\sigma^\epsilon_V = \lim_{k_z\rightarrow 0}  \frac i k_z \langle T^{0x}
T^{0y} \rangle \\
&&
 
\end{array}}
\end{equation}

All these correlators are to be taken at precisely zero frequency. As these
formulas are based on linear response theory
the correlators should be understood as retarded ones. They have to be evaluated
however at zero frequency and
therefore the order of the operators can be reversed. From this it follows that
the chiral vortical conductivity coincides
with the chiral magnetic conductivity for the energy flux $\sigma_V =
\sigma^\epsilon_B$.~\footnote{Notice that $h_{0y}$ can also
be understood as the so-called
gravito-magnetic vctor potential $\vec{A}_g$, which is related to the
gravito-magntic
 field by  ${\cal \vec{B}}_g  = \vec{\nabla}\times \vec{A}_g$. This allows to 
interpret $\sigma_V$
not only as the generation of a current due to a vortex in the fluid, i.e. the
chiral vortical effect, but also as a {\it  chiral gravito-magnetic}
conductivity giving rise to a {\it chiral gravito-magnetic effect},
see~\cite{Landsteiner:2011tg} for details. }

We also want to discuss how these transport coefficients are related to the ones
in the more commonly used Landau frame. 
They are connected by a redefinition of the fluid velocity of the form 
\begin{equation}
 u^\mu \rightarrow u^\mu - \frac{1}{\epsilon+p} Q^\mu \,,
\end{equation}
to go from (\ref{eq:alternativeframe1})-(\ref{eq:alternativeframe3}) to
(\ref{eq:Landauframe1})-(\ref{eq:Landauframe2}).
The corresponding transport coefficients of the Landau frame are therefore
\begin{eqnarray}
 \label{eq:cmc}\xi_{ B}  &=& \lim_{k_n\rightarrow 0} \frac{-i}{2 k_n} \sum_{k.l}
\epsilon_{nkl} \left( \left\langle J^{k} J^{l} \right\rangle - 
\frac{\rho}{\epsilon+p}\left\langle T^{0k} J^{l} \right\rangle \right)\,,\\
\label{eq:cvc}\xi_{V} &=& \lim_{k_n\rightarrow 0} \frac{-i}{2 k_n} \sum_{k.l}
\epsilon_{nkl}\left( \left\langle J^k T^{0l} \right\rangle - 
\frac{\rho}{\epsilon+p}\left\langle T^{0k} T^{0l} \right\rangle \right) \,,
\end{eqnarray}
where we have employed a slightly more covariant notation. The generalization to
the non-abelian case is straightforward.

It is also worth to compare to the Kubo formulae for the dissipative transport
coefficients as the electric conductivity (\ref{eq:sigmaelectric}).
In the dissipative cases one first goes to zero momentum and then takes the zero
frequency limit. In the anomalous conductivities this is
the other way around, one first goes to zero frequency and then takes the zero
momentum limit. Another observation is that the dissipative
transport coefficients sit in the anti-Hermitean part of the retarded
correlators, i.e. the spectral function whereas the anomalous 
conductivities sit in the Hermitean part. 
The rate at which an external source $f_I$ does work on a system is given in
terms of the spectral function of the operator $O^I$ coupling
to that source as
\begin{equation}
 \frac{dW}{dt} = \frac 1 2 \omega f_I(-\omega) \rho^{IJ}(\omega) f_J(\omega) \,.
\end{equation}
The anomalous transport phenomena therefore do no work on the system, first they
take place at zero frequency and second they are not
contained in the spectral function $\rho = \frac{-i}{2} (G_{r}-G_r^\dagger)$.

 \subsection{Contributions to the Kubo formulae}
 \label{subsec:contribs}
\begin{figure}
 \includegraphics[scale=.45]{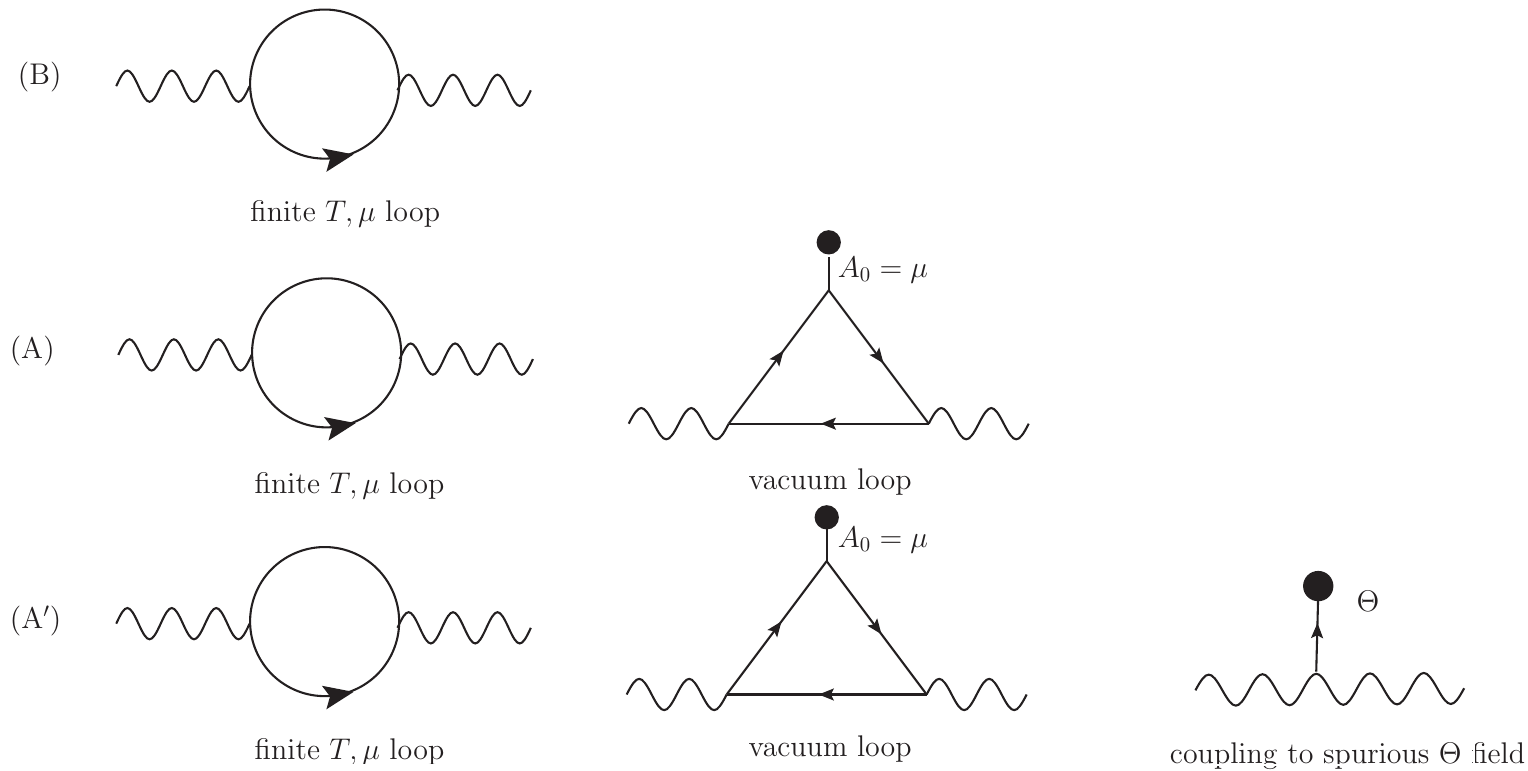}
 \caption{Contributions to the Kubo formula for the chiral magnetic
conductivity
 in the different formalisms for the chemical potential.}
 \label{fig:contribs}       
 \end{figure}

 Now we want to give a detailed analysis of the different Feynman
 graphs that contribute to the Kubo formulae in the different
 formalisms for the chemical potentials. The simplest and most economic
 formalism is certainly the one labeled (B) in which we introduce the
 chemical potentials via twisted boundary conditions. The Hamiltonian
 is simply the microscopic Hamiltonian $H$. Relevant contributions
 arise only at first order in the momentum and at zero frequency and in
 this kinematic limit only the Kubo formulae for the chiral magnetic
 conductivity is affected. In the figure (\ref{fig:contribs}) we
 summarize the different contributions to the Kubo formulae in the
 three ways to introduce the chemical potential.
 
 The first of the Feynman graphs is the same in all formalisms. It is the
genuine
 finite temperature and finite density one-loop contribution. This graph is
 finite because the Fermi-Dirac distributions cutoff the UV momentum modes in
the
 loop. In the formalism $(A)$ we need to take into account that there is
 also a contribution from the triangle graph with the fermions going around the
 loop in vacuum, i.e. without the Fermi-Dirac distributions in the loop
 integrals.
 For a non-anomalous symmetry this graph vanishes simply because on the upper
 vertex of the triangle sits a field configuration that is a pure gauge. If the
 symmetry
 under consideration is however anomalous the triangle diagram picks up just the
 anomaly. Even pure gauge field configurations become physically distinct from
 the
 vacuum and therefore this diagram gives a non-trivial contribution. On the
level
 of the constitutive relations this contribution corresponds to the Chern-Simons
 current in (\ref{eq:consistentconstitutive}).
 We consider this contribution to be unwanted. After all the anomaly would make
 even a constant
 value of the temporal gauge field $A_0$ observable in vacuum. An example is
 provided for a putative axial gauge field $A^5_\mu$. If present the absolute
 value of
 its temporal component would be observable through the axial anomaly. We can be
 sure that in nature no such background field is present. 
 The third line $(A')$ introduces also the spurious axion field $\Theta$ the
only
 purpose of this field is to cancel the contribution from the triangle graph. 
 This cancellation takes place by construction since $(A')$ is gauge equivalent
 to $(B)$ in which only the first genuine finite $T,\mu$ part
 contributes. It corresponds to the contribution of the current $J^\mu_\Theta$
in (\ref{eq:thetaconstitutive}).We further emphasize that these considerations 
are based on the usage of the consistent currents. 

In the interplay between axial and vector
 currents additional contributions arise from the Bardeen
 counterterm. It turns out that the triangle or Chern-Simons current
 contribution to the consistent vector current in the formalism $(A)$
 cancels precisely the first one
 \cite{Rebhan:2009vc,Gynther:2010ed}. Our take on this is that a
 constant temporal component of the axial gauge field $A_0=\mu_5$ would
 be observable in nature and can therefore be assumed to be absent. The
 correct way of evaluating the Kubo formulae for the chiral magnetic
 effect is therefore the formalism $(B)$ or the gauge equivalent one
 $(A')$.
 
 At this point the reader might wonder why we introduced yet another formalism
$(A')$
 which achieves appearently nothing but being equivalent to formalism $(B)$. At
least
 from the perspective of holography there is a good reason for doing so. 
 In holography the strong coupling duals of gauge theories
 at finite temperature in the plasma phase are represented by five dimensional
 asymptotically Anti- de Sitter black holes. Finite charge density translates to
 charged black holes. These black holes have some non-trivial gauge flux along
 the holographic direction represented by a temporal gauge field configuration
 of the form $A_0(r)$ where $r$ is the fifth, holographic dimension. It is often
 claimed that for consistency reasons the gauge field has to vanish on the 
 horizon of the black hole and that its value on the boundary can be identified
 with the chemical potential
 \begin{equation}\label{eq:horizonzero}
  A_0(r_H) =0   ~~~~\mathrm{and}~~~ A_0(r\rightarrow \infty) = \mu\,.
 \end{equation}
 According to the usual holographic dictionary the gauge field values on the
 boundary correspond to the sources for currents. A non-vanishing
 value of the temporal component of the gauge field at the boundary is therefore
 dual to a coupling that modifies the Hamiltonian of the theory just as in
 (\ref{eq:deformedH}). Thus with the boundary conditions (\ref{eq:horizonzero})
 we have the holographic dual of the formalism $(A)$. 
 If anomalies are present they are represented in the holographic dual by
 five-dimensional Chern-Simons terms of the form $A\wedge F\wedge F$. The two
 point correlator
 of the (consistent) currents receives now contributions from the Chern-Simons
 term that is precisely of the form of the second graph in $(A)$ in figure
 \ref{fig:contribs}.
 As we have argued this is an a priory unwanted contribution. We can however
cure
 that by introducing an additional term in the action of the form
 (\ref{eq:Stheta})
 living only on the boundary of the holographic space-time. In this way we can
 implement the formalism $(A')$, cancel the unwanted triangle contribution
 with the third graph in $(A')$ in figure \ref{fig:contribs} and maintain
$A_0(r_H)=0$!
 
 The claim that the temporal component of the gauge field has to vanish at the
 horizon is of course not unsubstantiated. The reasoning goes as follows. The
 Euclidean section
 of the black-hole space time has the topology of a disc in the $r,\tau$
 directions, where $\tau$ is the Euclidean time. This is a periodic variable
with
 period $\beta=1/T$ 
 where $T$ is the (Hawking) temperature of the black hole and at the same time
the
 temperature in the dual field theory. Using Stoke's law we have
 \begin{equation}
  \int_{\partial D} A_0 \,d\tau = \int_D F_{r0}\,dr\, d\tau \,,
 \end{equation}
 where $F_{r0}$ is the electric field strength in the holographic
 direction and $D$ is a Disc with origin at $r=r_H$ reaching out to
 some finite value of $r_f$.  If we shrink this disc to zero size,
 i.e. let $r_f\rightarrow r_H$ the r.h.s. of the last equation vanishes
 and so must the l.h.s. which approaches the value $ \beta A_0(r_H)$.
 This implies that $A_0(r_H)=0$.  If on the other hand we assume that
 $A_0(r_H)\neq 0$ then the field strength must have a delta type
 singularity there in order to satisfy Stokes theorem.  Strictly
 speaking the topology of the Euclidean section of the black hole is
 not anymore that of a disc since now there is a puncture at the
 horizon. It is therefore more appropriate to think of this as having
 the topology of a cylinder.  Now if we want to implement the formalism
 $(B)$ in holography we would find the boundary conditions
 \begin{equation}\label{eq:boundaryzero}
  A_0(r_H) = \mu   ~~~~\mathrm{and}~~~ A_0(r\rightarrow \infty) = 0\,,
 \end{equation}
 and precisely such a singularity at the horizon would arise. In addition we
 would need to impose twisted boundary conditions around the Euclidean time
 $\tau$ for the fields
 just as in (\ref{eq:bcs2}). Now the presence of the singularity seems to be a
 good thing: if the space time would still be smooth at the horizon it would be
 impossible
 to demand these twisted boundary conditions since the circle in $\tau$ shrinks
 to zero size there. If this is however a singular point of the geometry we can
 not
 really shrink the circle to zero size. The topology being rather a cylinder
than
 a disc allows now for the presence of the twisted boundary conditions. 
 
 It is also important to note that in all formalisms the potential difference
 between the boundary and the horizon is given by $\mu$. This has a very nice
 intuitive 
 interpretation. If we bring a unit test charge from the boundary to the horizon
 we need the energy $\Delta E = \mu$. In the dual field theory this is just the
 energy cost
 of adding one unit of charge to the thermalized system and coincides with the
 elementary definition of the chemical potential. 
 \begin{figure}

 \scalebox{0.4}{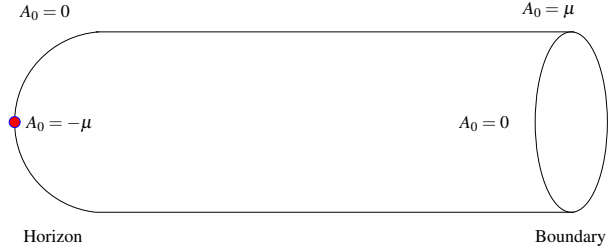 }

 \caption{A sketch of the Euclidean black hole topology. A singularity at the
 horizon arises if we do not choose the temporal component of the gauge field to
 vanish there.
 On the other hand allowing the singularity to be present changes the topology
to
 the one of a cylinder and this in turn allows twisted boundary conditions.}
 \label{fig:EuclideanBH}      
 \end{figure}
 
 From now on we will always only consider the genuine finite $T,\mu$
contribution
 that is the only one that arises in formalism $(B)$. 
 
 The rest of this review is devoted to the explicit evaluation of these Kubo
 formulae in two different systems: free chiral fermions and
 a holographic model implementing the chiral and gravitational anomalies by
 suitable five dimensional Chern-Simons terms.

\section{Weyl fermions}
\label{sec:3}

We will now evaluate the Kubo formulae for the chiral magnetic, chiral
vortical and energy flux conductivities (\ref{Kformulae}) for a theory
of $N$ free chiral fermions $\Psi^f$ transforming under a global
symmetry group $G$ generated by matrices $(T_a)^f\,_g$.  

We denote the generators in the Cartan subalgebra by $H_a$. Chemical
potentials $\mu_a$ can be switched on only in the Cartan
subalgebra. Furthermore the presence of the chemical potentials breaks
the group $G$ to a subgroup $\hat G$. Only the currents that lie in
the unbroken subgroup are conserved (up to anomalies) and participate
in the hydrodynamics. The chemical potential for the fermion $\Psi^f$
is given by $\mu^f= \sum_a q_a^f \mu_a$, where we write the Cartan
generator $H_a = q_a^f\delta^f\,_g$ in terms of its eigenvalues, the
charges $q_a^f$. The unbroken symmetry group $\hat G$ is generated by
those matrices $T_a^f\,_g$ fulfilling
\begin{equation}\label{eq:unbroken}
 T_a^f\,_g \mu^g = \mu^f T_a^f\,_g\,.
\end{equation}
There is no summation over indices in the last expression. From now on we will
assume that all currents $\vec{J}_a$ lie in directions indicated in 
(\ref{eq:unbroken}). We define the chemical potential through the boundary
condition on the fermion fields around the thermal circle, i.e. we adopt the
formalism (B) discussed in previous section,
\begin{equation}
 \Psi^f(\tau-\beta) = - e^{\beta\mu^f} \Psi^f(\tau) \,.
\end{equation}
Therefore the eigenvalues of $\partial_\tau$ are
$i\tilde\omega_n+\mu^f$ for the fermion spiecies $f$ with
$\tilde\omega_n=\pi T(2n+1)$ the fermionic Matsubara frequencies.  A
convenient way of expressing the current and the energy-momentum tensor is in
terms of Dirac fermions and writing
\begin{equation}
J^i_a = \sum_{f,g=1}^N T_a^g\,_f \bar\Psi_g \gamma^i \cP_+ \Psi^f \,, \qquad
T^{0i} =  \frac i 2 \sum_{f=1}^N\bar\Psi_f  ( \gamma^0  \partial^i + \gamma^i
\partial^0  ) \cP_+\Psi^f\,, \label{eq:JAJE}
\end{equation}
where we used the chiral projector $\cP_\pm = \frac 1 2 (1\pm\gamma_5)$.
The fermion propagator is
\begin{equation}
S(q)^f\,_g =  \frac{\delta^f\,_g}{2} \sum_{t=\pm}
\Delta_t(i\tilde\omega^f,\vec{q}) \cP_+ \gamma_\mu \hat q^\mu_t \,, \qquad
\Delta_t( i\tilde\omega^f, q) = \frac{1}{i\tilde\omega^f - t E_q}\,,
\end{equation}
with  $i\tilde\omega^f = i\tilde\omega_n + \mu^f$, $\hat q_t^\mu = (1, t \hat
q)$, $\hat{q} = \frac{\vec{q}}{E_q}$ and $E_q=|\vec q |$. For simplicity in the
expressions we consider only left-handed fermions, but one can easily include
right-handed fermions as well as they contribute in all our calculations in the
same way as the left-handed ones up to a relative minus sign.

We will address in detail the computation of the vortical
conductivities and sketch only the calculation of the magnetic
conductivities since the latter one is a trivial extension of the
calculation of the chiral magnetic conductivity in
\cite{Kharzeev:2009pj}. Then we show the results for the other
conductivity coefficients.

\subsection{Chiral Vortical Conductivity}
\label{subsec:vortical}

\begin{figure}[tbp]
\begin{center}
\includegraphics[scale=1]{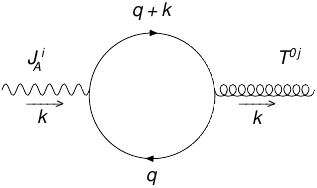}
\end{center}
\caption{1 loop diagram contributing to the vortical conductivity
eq.~(\ref{eq:Gavx}).}
\label{fig:1loop}
\end{figure}

 The vortical conductivity is defined from the retarded correlation
 function of the current $J^i_a(x)$ and the energy momentum tensor or
 energy current $T^{0j}(x^\prime)$~(\ref{eq:JAJE}), i.e.
\begin{equation}
G_a^V(x-x^\prime) =  \frac{1}{2} \epsilon_{ijn}\,i \, \theta(t-t^\prime) 
\,\langle [J^i_a(x),T^{0j}(x^\prime)] \rangle \,. \label{eq:Gavx}
\end{equation}
Going to Fourier space, one can evaluate this quantity as
\begin{equation}\label{eq:Gav}
 G_a^V(k) = \frac{1}{4}\sum_{f=1}^N T_a^f\,_f
\frac{1}{\beta}\sum_{\tilde\omega^f} \int\frac{d^3q}{(2\pi)^3}
\scriptstyle{\epsilon_{ijn} \tr \Big[ S^f\,_f(q) \gamma^i 
\,S^f\,_f(q+k) \left(  \gamma^0 q^j + \gamma^j i\tilde\omega^f \right)  \Big]}
\, ,
\end{equation}
which corresponds to the one loop diagram of figure~\ref{fig:1loop}. 
The vertex of the two quarks with the graviton is $\sim \delta^f\,_g$, and
therefore we find only contributions from the diagonal part of the group~$\hat
G$. 
The metric we use through this section is the usual one in field theory
computations, $g_{\mu\nu} = \mathrm{diag}(1,-1,-1,-1)$. We can split $G_a^V$
into two contributions, i.e.
\begin{equation}
G_a^V(k) = G_{a,(0j)}^V(k) + G_{a,(j0)}^V(k) \,, \label{eq:Gav2}
\end{equation}
which correspond to the terms $\gamma^0 q^j$  and $\gamma^j i\tilde\omega^f$ in
eq.~(\ref{eq:Gav}) respectively. We will focus first on the computation of
$G_{a,(0j)}^V$. After computation of the Dirac trace in eq.~(\ref{eq:Gav}), this
term writes
\begin{eqnarray}
G_{a,(0j)}^V(k) &=& \frac{1}{8} \sum_{f=1}^N  T_a^f\,_f
\frac{1}{\beta}\sum_{\tilde\omega^f} \int \frac{d^3q}{(2\pi)^3}   q^j
\sum_{t,u=\pm} \Bigg[  \epsilon_{ijn} \Bigg( t\frac{q^i}{E_q}+u \frac{k^i +
q^i}{E_{q+k}}\Bigg) \nonumber \\
&&\; + i\frac{tu}{E_q E_{q+k}} \left( q_j k_n - q_n k_j  \right) \Bigg]  
\Delta_t(i\tilde\omega^f,\vec{k})
\Delta_u(i\tilde\omega^f+i\omega_n,\vec{q}+\vec{k})  \,. \label{eq:Gav0j2}
\end{eqnarray}
At this point one can make a few simplifications. Note that due to the
antisymmetric tensor $\epsilon_{ijn}$, the two terms proportional to $q^i$
inside the bracket in eq.~(\ref{eq:Gav0j2}) vanish. Regarding the term 
$\epsilon_{ijn} q^j k^i$, it leads to a contribution $\sim \epsilon_{ijn} k^j
k^i$ after integration in $d^3q$, which is zero. Then the only term which
remains is the one not involving $\epsilon_{ijn}$. We can now perform the sum
over fermionic Matsubara frequencies. One has
\begin{equation} \label{eq:Matsubara1}
\frac{1}{\beta}\sum_{\tilde\omega^f}
{\textstyle\Delta_t(i\tilde\omega^f,\vec{q})
\Delta_u(i\tilde\omega^f+i\omega_n,\vec{q}+\vec{p}) = \frac{t n(E_q - t\mu^f) -
u n(E_{q+k}-u \mu^f) + \frac{1}{2}(u-t)}{i\omega_n + t E_q - u E_{q+k}}} \,,
\end{equation}
where $n(x) = 1/(e^{\beta x} + 1)$ is the Fermi-Dirac distribution function. In
eq.~(\ref{eq:Matsubara1}) we have considered that $\omega_n = 2\pi T n$ is a
bosonic Matsubara frequency. This result is also obtained in
Ref.~\cite{Kharzeev:2009pj}. After doing the analytic continuation, which
amounts to replacing $i\omega_n$ by $k_0 + i\epsilon$ in
eq.~(\ref{eq:Matsubara1}), one gets
\begin{eqnarray}
G_{a,(0j)}^V(k) &=& -\frac{i}{8} \sum_{f=1}^N T_a^f\,_f \int
\frac{d^3q}{(2\pi)^3}  \frac{ \vec{q}^2 k_n - (\vec{q}\cdot\vec{k}) q_n }{E_q
E_{q+k}} \times \nonumber \\
&&\qquad\qquad \sum_{t,u=\pm}  \frac{u n(E_q - t\mu^f) - t n(E_{q+k}-u \mu^f) +
\frac{1}{2}(t-u)}{k_0 + i\epsilon + t E_q - u E_{q+k}}\,. \label{eq:Gav0j3}
\end{eqnarray} 
The term proportional to~$\sim \frac{1}{2}(t-u)$ corresponds to the vacuum
contribution, and it is ultraviolet divergent. By removing this term the finite
temperature and chemical potential behavior is not affected, and the result
becomes ultraviolet finite because the Fermi-Dirac distribution function
exponentially suppresses high momenta. By making both the change of variable
$\vec{q} \to -\vec{q} -\vec{k}$ and the interchange $u \to -t$ and $t \to -u$ in
the part of the integrand involving the term $-tn(E_{q+k}-u\mu^f)$, one can
express the vacuum substracted contribution of eq.~(\ref{eq:Gav0j3}) as
\begin{eqnarray}
\widehat{G}_{a,(0j)}^V(k) &=& \frac{i}{8} k_n \sum_{f=1}^N T_a^f\,_f \int
\frac{d^3 q}{(2\pi)^3} \frac{1}{E_q E_{q+k}} \left( \vec{q}^2 -
\frac{(\vec{q}\cdot\vec{k})^2}{\vec{k}^2}\right) \times \nonumber \\
&&\qquad\qquad \sum_{t,u=\pm} u \frac{n(E_q - \mu^f) + n(E_q+ \mu^f)}{k_0 +
i\epsilon + t E_q + u E_{q+k}}\,, \label{eq:Gav0j4}
\end{eqnarray}
where we have used that $n(E_q - t\mu^f) + n(E_q+ t\mu^f) = n(E_q - \mu^f) +
n(E_q+ \mu^f)$ since $t=\pm 1$. The result has to be proportional to $k_n$, so
to reach this expression we have replaced $q_n$ by $(\vec{q}\cdot
\vec{k})k_n/\vec{k}^2$ in eq.~(\ref{eq:Gav0j3}). At this point one can perform
the sum over $u$ by using $\sum_{u = \pm} u/(a_1+u a_2) = -2a_2/(a_1^2-a_2^2)$,
and the integration over angles by considering $\vec{q} \cdot \vec{k} = E_q E_k
x $ and $E_{q+k}^2 = E_q^2+E_k^2 + 2 E_q E_k x$, where $x := \cos(\theta)$ and
$\theta$ is the angle between $\vec{q}$ and $\vec{k}$. Then one gets the final
result
\begin{eqnarray}
\widehat{G}_{a,(0j)}^V(k) &=& \frac{i}{16\pi^2}\frac{k_n}{k^2}(k^2 - k_0^2)
\sum_{f=1}^N T_a^f\,_f \int_0^\infty dq \, q \, f^V(q) \times \label{eq:Gav0j5}
\\
&&\qquad \Bigg[ 1 + \frac{1}{8qk} \sum_{t=\pm} \left[k_0^2-k^2+4q(q+t
k_0)\right] \log \left( \frac{\Omega_t^2 - (q+k)^2}{\Omega_t^2 - (q-k)^2}\right)
\Bigg] \,,\nonumber
\end{eqnarray} 
where $\Omega_t = k_0 + i\epsilon + t E_q \,,$ and
\begin{equation}
f^V(q) = n(E_q-\mu^f) + n(E_q + \mu^f) \,.   \label{eq:fV}
\end{equation}
The steps to compute $G_{a,(j0)}^V$ in eq.~(\ref{eq:Gav2}) are similar. In this
case the Dirac trace leads to a different tensor structure, in  which the only
contribution comes from the trace involving~$\gamma_5$. The sum over fermionic
Matsubara frequencies involves an extra~$i\tilde\omega^f$, i.e.
\begin{eqnarray}
&&\frac{1}{\beta}\sum_{\tilde\omega^f} i{\tilde\omega^f}
\Delta_t(i\tilde\omega^f,\vec{q})
\Delta_u(i\tilde\omega^f+i\omega_n,\vec{q}+\vec{k}) = \frac{1}{i\omega_n + t E_q
- u E_{q+k}} \times \label{eq:Matsubara2} \\
&&\qquad \Bigg[ E_q n(E_q - t\mu^f)  - (E_{q+k} - ui\omega_n) n(E_{q+k} -u
\mu^f) -\frac{1}{2}\left( E_q - E_{q+k} + ui\omega_n \right) \Bigg] \,.
\nonumber
\end{eqnarray}
The last term inside the bracket in the r.h.s. of eq.~(\ref{eq:Matsubara2})
corresponds to the vacuum contribution which we choose to remove, as it leads to
an ultraviolet divergent contribution after integration in $d^3q$. Making
similar steps as for $\widehat{G}_{a,(0j)}^V$, i.e. performing the sum over $u$
and integrating over angles, one gets the final result
\begin{eqnarray}
\widehat{G}_{a,(j0)}^V(k) &=& -\frac{i}{32\pi^2} \frac{k_n}{k^3}\sum_{f=1}^N
T_a^f\,_f \int_0^\infty dq \sum_{t=\pm} f^V_t(q,k_0) \times \nonumber \\
 &&\qquad \Bigg[  4 t q k k_0  -\left(k^2-k_0^2\right) (2q + t k_0) \log\left(
\frac{\Omega_t^2  - (q+k)^2}{\Omega_t^2 - (q-k)^2 } \right) \Bigg]
\,,\label{eq:Gavj05} 
\end{eqnarray}
where
\begin{equation}
f^V_t(q,k_0) = q f^V(q) + t k_0 n(E_q + t\mu^f)   \,. \label{eq:fVt}
\end{equation}
The result for the vacuum substracted contribution of the retarded correlation
function of the current and the energy momentum tensor, $\hat{G}_a^V(k)$, writes
as a sum of eqs.~(\ref{eq:Gav0j5}) and (\ref{eq:Gavj05}), according to
eq.~(\ref{eq:Gav2}). From these expressions one can compute the zero frequency,
zero momentum, limit. Since
\begin{equation}
\lim_{k\to 0} \lim_{k_0 \to 0} \sum_{t=\pm}\log \left( \frac{\Omega_t^2 -
(q+k)^2}{\Omega_t^2 - (q-k)^2}\right) = \frac{2k}{q} \,, \label{eq:limlog}
\end{equation}
the relevant integrals are
\begin{equation}
\int_0^\infty dq \, q \, f^V(q) = \int_0^\infty dq \, f^V_t(q,k_0 = 0) =
\frac{(\mu^f)^2}{2} + \frac{\pi^2}{6} T^2 \,.
\end{equation}
Finally it follows from eqs.~(\ref{eq:Gav0j5}) and (\ref{eq:Gavj05}) that the
zero frequency, zero momentum, vortical conductivity writes
\begin{eqnarray}
( \sigma_V )_a &=& \frac{1}{8\pi^2} \sum_{f=1}^N T_a^f\,_f \Bigg[ ( \mu^f)^2 +
\frac{\pi^2}{3} T^2 \Bigg]  \nonumber \\
&=&\frac{1}{16\pi^2}  \Bigg[ \sum_{b,c} \tr\left(T_a \{ H_b , H_c \}\right)
\mu_b \, \mu_c  + \frac{2\pi^2}{3} T^2 \tr \left( T_a \right) \Bigg]  \,.
\label{eq:vorticalconductivity}
\end{eqnarray}
Both $\widehat{G}_{a,(0j)}^V$ and $\widehat{G}_{a,(j0)}^V$ lead to the same
contribution in~$(\sigma_V)_a$. Eq.~(\ref{eq:vorticalconductivity}) was first
derived in~\cite{Landsteiner:2011cp}, and it constitutes our main result in this
section, . The term involving the chemical potentials is induced by the chiral
anomaly. More interesting is the term $\sim T^2$ which is proportional to the
gravitational anomaly coefficient
$b_a$~\cite{Delbourgo:1972xb,Delbourgo2, Eguchi, AlvarezGaume:1983ig}. This means
that a non-zero value of this term has to be attributed to the presence of a
gravitational anomaly. The Matsubara frequencies $\tilde\omega_n=\pi T(2n+1)$
generate a dependence on $\pi T$ in the final result as compared to the chemical
potentials, and then no factors of $\pi$ show up for the term $\sim T^2$ in
eq.~(\ref{eq:vorticalconductivity}). Right-handed fermions contribute in the
same way but with a relative minus sign. Therefore the $\sim T^2$ term appears
only when the current in eq.~(\ref{eq:Gavx}) has an axial component. The
correlator with a vector current does not have this gravitational anomaly
contribution.

\subsection{Chiral Magnetic Conductivity}
\label{sec:magnetic-conductivity}

The chiral magnetic conductivity in the case of a vector and an axial $U(1)$
symmetry was computed at weak coupling in~\cite{Kharzeev:2009pj}. The
corresponding Kubo formula involves the two point function of the current, see
first expression in eq.~(\ref{Kformulae}). Following the same method, we have
computed it for the unbroken (non-abelian) symmetry group~$\hat G$.
The relevant Green function is~\cite{Landsteiner:2011cp} 
\begin{eqnarray}
G^B_{ab}(k) = \frac 1 2 \sum_{f,g} T_a^f\,_g T_b^g\,_f \frac 1 \beta 
\sum_{\tilde\omega^f} \int\frac{d^3q}{(2\pi)^3} \epsilon_{ijn} \tr \Bigg[
S^f\,_f(q) \gamma^i S^f\,_f(q+k) \gamma^j    \Bigg] \,. \label{eq:Gabb}
\end{eqnarray}
The evaluation of this expression is exactly as in \cite{Kharzeev:2009pj} so we
skip the details. The zero frequency, zero momentum, limit of the magnetic
conductivity is 
\begin{equation}
(\sigma_B)_{ab} =  \frac{1}{4\pi^2} \sum_{f,g=1}^N T_a^f\,_g T_b^g\,_f \, \mu^f
= \frac{1}{8\pi^2} \sum_c \tr\left( T_a \{ T_b , H_c \} \right) \, \mu_c \,.
\label{eq:sigmaB1}
\end{equation}
In the second equality of eq.~(\ref{eq:sigmaB1}) we have made use of
eq.~(\ref{eq:unbroken}). No contribution proportional to the
gravitational anomaly coefficient is found in this case.

\subsection{Conductivities for the Energy Flux}
\label{sec:energy_flux} 

We will include for completeness the result of the chiral magnetic and vortical
conductivities for the energy flux, corresponding to the last two expressions in
eq.~(\ref{Kformulae}).
  
The chiral magnetic conductivity for energy flux, $\sigma_B^{\epsilon}$, follows
from the correlation function of the energy momentum tensor and the current, and
so it computes in the same way as the vortical conductivity in
Sec.~\ref{subsec:vortical}. From an evaluation of the corresponding Feynman
diagram one finds that the result is the same as eq.~(\ref{eq:Gav}). Then one
concludes that
\begin{equation}
(\sigma_B^{\epsilon})_a = (\sigma_V)_a \,, \label{eq:sigmaeB}
\end{equation}
where $(\sigma_V)_a$ is given by eq.~(\ref{eq:vorticalconductivity}). Although
these coefficients are equal, they describe different transport phenomena.
Whereas~$(\sigma_B^{\epsilon})_a$ describes the generation of an energy flux due
to an external magnetic field~$\vec{B}_a$, $(\sigma_V)_a$ describes the
generation of the current~$\vec{J}_a$ due to an external field that sources the
energy-momentum tensor~$T^{0i}$. 

Finally the chiral vortical conductivity for the energy flux, $\sigma_V^{\epsilon}$, follows from the correlation function of two energy momentum tensors. In this case there is a contribution corresponding to the seagull diagram, which leads to the term
\begin{equation}\label{eq:Gseagull}
 G^\epsilon_V(k) |_{\textrm{seagull}} = \frac{1}{16}\sum_{f=1}^N T_a^f\,_f
\frac{1}{\beta}\sum_{\tilde\omega^f} \int\frac{d^3q}{(2\pi)^3}
\epsilon_{ijn} \tr \Big[  S(q) \left\{ \gamma^i \gamma^j , k\!\!\!\!/ \right\}  \Big]
\, .
\end{equation}
This was computed in~\cite{Manes:2012hf}. In the computation of the Green function for the chiral vortical conductivity of energy flux there are three contributions out of the four possible terms. One of these terms involves a sum over fermionic Matsubara frequencies of the
form
 \begin{eqnarray}
 &&\frac{1}{\beta}\sum_{\tilde\omega^f} (i{\tilde\omega^f})^2
\Delta_t(i\tilde\omega^f,\vec{q})
\Delta_u(i\tilde\omega^f+i\omega_n,\vec{q}+\vec{k})  =  {\cal
F}(i\omega_n,E_q,E_{q+k},t,u) + \label{eq:Matsubara3}  \\
&&\qquad + \frac{1}{i\omega_n + t E_q - u E_{q+k}} \Bigg[ t E_q^2  n(E_q -
t\mu^f) - u (E_{q+k} - ui\omega_n)^2 n(E_{q+k} -u \mu^f)  \Bigg]  \,, \nonumber
 \end{eqnarray}
 where $\cal F$ corresponds to the ultraviolet divergent vacuum contribution
which we choose to remove. The zero frequency, zero momentum, limit of the
chiral vortical conductivity for the energy flux writes
 \begin{eqnarray}
 \sigma_V^{\epsilon} &=& \frac{1}{12\pi^2} \sum_{f=1}^N \bigg[ (\mu^f)^3 + \pi^2
T^2 \mu^f \bigg] \nonumber \\
 &=& \frac{1}{24\pi^2}  \Bigg[ \sum_{a,b,c} \tr\left(H_a \{ H_b , H_c \}\right)
\mu_a \mu_b \, \mu_c  + 2\pi^2 T^2 \sum_a \tr\left(H_a\right) \mu_a \Bigg] \,. 
\label{eq:sigmaeV}
\end{eqnarray}
This coefficient describes the generation of an energy flux due to a vortex (or
a gravito-magnetic field). The correlators~(\ref{eq:sigmaeB}) and
(\ref{eq:sigmaeV}) enter the chiral magnetic and vortical conductivities in the
Landau frame, respectively, as defined
in~\cite{Erdmenger:2008rm,Banerjee:2008th,Son:2009tf}, see
eqs.~(\ref{eq:cmc})-(\ref{eq:cvc}). We have also checked that to lowest order in
$\omega$ and $k$ one has $\langle T^{0z} T^{0z} \rangle = p$, where $p$ is the
pressure of a free gas of massless fermions, and $\langle  T^{0z} J^z
\rangle=0$~\cite{Amado:2011zx}.

\subsection{Summary and specialization to the group $U(1)_V \times U(1)_A$}
\label{sec:VAcase}

The results for the different conductivities are neatly summarized as
\begin{eqnarray}
(\sigma_B)_{ab} &=& \frac{1}{4 \pi^2} d_{abc}\mu^c\,, \label{eq:cmclab} \\
(\sigma_V)_a &=&  (\sigma_B^\epsilon)_a = \frac{1}{8\pi^2} d_{abc} \mu^b\mu^c  +
\frac{T^2}{24} b_a\,, \label{eq:cvclab} \\
\sigma^\epsilon_V &=& \frac{1}{12\pi^2} d_{abc}\mu^a\mu^b\mu^c + \frac{T^2}{12}
b_a\mu^a\,.  \label{eq:cvcelab}
\end{eqnarray}
The axial and mixed gauge-gravitational anomaly coefficients are defined by
\begin{eqnarray}
d_{abc} &=& \frac 1 2 [ \tr( T_a \{ T_b, T_c\} )_L -  \tr( T_a \{ T_b, T_c\} )_R
]\,, \\
b_a &=& \tr (T_a)_L - \tr (T_a)_R\,,
\end{eqnarray}
where the subscripts $L,R$ stand for the contributions of left-handed and
right-handed fermions. The result shows that these conductivities are non-zero
if and only if the theory features anomalies.

For phenomenological reasons  it is interesting to specialize
these results to the symmetry group $U(1)_V \times U(1)_A$, i.e. one vector
and one axial current with chemical potentials $\mu_L = \mu+\mu_A$,
$\mu_R= \mu -\mu_A$, charges $q^L_{V,A}=(1,1)$ and $q^R_{V,A}=(1,-1)$
for one left-handed and one right-handed fermion. We find (for a
vector magnetic field)
\begin{eqnarray}
&(\sigma_B)_{VV}& = \frac{\mu_A}{2\pi^2}  \,, \qquad\qquad\qquad\qquad
(\sigma_B)_{AV} =
\frac{\mu}{2\pi^2} \,,  \\
&(\sigma_V)_{V}&= (\sigma_B^\epsilon)_V = \frac{\mu\mu_A}{2\pi^2} \,,
\qquad\qquad
(\sigma_V)_{A} = (\sigma_B^\epsilon)_A =
\frac{\mu^2+\mu_A^2}{4\pi^2}+ \frac{T^2}{12}   \,,\\
&\sigma_V^\epsilon& =   \frac{\mu_A}{6\pi^2} \left( 3\mu^2 + \mu_A^2 \right) +
\frac{\mu_A}{6}T^2 \,. \label{eq:axialcvc}
\end{eqnarray}
Here $(\sigma_B)_{VV}$ is the chiral magnetic conductivity
\cite{Kharzeev:2009pj}, $(\sigma_B)_{AV}$ describes the generation of
an axial current due to a vector magnetic field~\cite{Son:2004tq},
$(\sigma_V)_V$ is the vector vortical conductivity, $(\sigma_V)_A$ is the axial
vortical conductivity, and $\sigma_V^\epsilon$ is the vortical conductivity for
the energy flux. The vector and axial magnetic conductivities for energy flux
$(\sigma_B^\epsilon)_V$ and $(\sigma_B^\epsilon)_A$ coincide with the chiral
vortical conductivities.

\section{Holographic Model}
\label{sec:4}
In this section for simplicity we will consider a holographic system which
realize a single chiral $U(1)$ symmetry with a gauge and mixed
gauge-gravitational
anomaly \cite{Landsteiner:2011iq}. As we saw in the previous section in a more
realistic model $U(1)_V\times U(1)_A$ the transport coefficients receive
contribution from the gravitational part only in the axial sector. For a study
of such a system with a pure gauge anomaly using Kubo formulae,
see~\cite{Gynther:2010ed}.

\subsection{Notation and Holographic Anomalies}

Let us fix some conventions we will use in the Gravity Theory. We choose the
five dimensional metric to be of signature $(-,+,+,+,+)$.  Five dimensional
indices are denoted with upper case latin letters. The epsilon tensor has to be
distinguished from the epsilon symbol by
$\epsilon_{ABCDE}=\sqrt{-g}\,\epsilon(ABCDE)$. The symbol is defined by
$\epsilon(rtxyz) = +1$. We assume the metric can be decomposed in ADM like way
and define an outward pointing normal vector to the holographic boundary of an
asymptotically $AdS$ space $n_A \propto g^{AB} \frac{\partial r}{\partial x^B}$
with unit norm $n_A n^A =1$.   So that the induced metric takes the form 
\begin{equation}\label{eq:inducemetric}
 h_{AB} = g_{AB}- n_A n_B\,.
\end{equation}

In general a foliation of the space-time $M$ with timelike surfaces defined
through $r(x) = \mathrm{const}$ can be written as
\begin{equation}
 ds^2 = (N^2 + N_A N^A) dr^2 + 2N_A dx^A dr+ h_{AB}dx^A dx^B\,.
\end{equation}

The Christoffel symbols,  Riemann tensor and extrinsic curvature are given by
\begin{eqnarray}
 \Gamma^M_{NP} &=& \frac{ 1}{ 2} g^{MK}\left(  \partial_N g_{KP} +  \partial_P
g_{KM} - \partial_K g_{NP}  \right),\\ 
 R^M\,_{NPQ} &=& \partial_P \Gamma^M_{NQ} - \partial_Q \Gamma^M_{NP} +
\Gamma^M_{PK} \Gamma^K_{NQ} -   \Gamma^M_{QK} \Gamma^K_{NP} ,\\
 K_{AV} &=&   h_A^C \nabla_C n_V =  \frac 1 2 {\pounds}_n h_{AB} \,,
\end{eqnarray}
where $\pounds_n$ denotes the Lie derivative in direction of $n_A$. Finally we
can define our model. The action is given by 
\begin{eqnarray}
\nonumber
S &=& \frac{1}{16\pi G} \int_M d^5x \sqrt{-g} \left[ R + 2 \Lambda -
 \frac 1 4 F_{MN} F^{MN} \right. \\
&&\left.+ \epsilon^{MNPQR} A_M
\left( \frac\kappa 3 F_{NP} F_{QR} + \lambda R^A\,_{BNP} R^B\,_{AQR}
\right) \right] + S_{GH} + S_{CSK} \,, 
\end{eqnarray}
\begin{eqnarray}
S_{GH} &=& \frac{1}{8\pi G}
\int_{\partial M} d^4x \sqrt{-h} \, K \,,\\ 
S_{CSK} &=& - \frac{1}{2\pi G}
\int_{\partial M} d^4x \sqrt{-h} \, \lambda n_M \epsilon^{MNPQR} A_N
K_{PL} D_Q K_R^L \,, 
\end{eqnarray}
where $S_{GH}$ is the usual Gibbons-Hawking boundary term and $D_A$ is
the induced covariant derivative on the four dimensional surface. The
second boundary term $S_{CSK}$ is introduced to reproduce the
gravitational anomaly at general hypersurface.  

Lets study now the gauge symmetries of our model. We note that the
action is diffeomorphism invariant, but they do depend however
explicitly on the gauge connection $A_M$.  Under gauge transformations
$\delta A_M = \nabla_M \xi$ they are therefore invariant only up to a
boundary term. We have
\begin{eqnarray}
 \delta S &=& \frac{1}{16\pi G} \int_{\partial M} d^4x \sqrt{-h} \,\xi 
\epsilon^{MNPQR} \left( \frac{\kappa}{3}n_M F_{NP}F_{QR} +
\lambda n_MR^A\,_{BNP} R^{B}\,_{AQR}\right) -\,\nonumber\\
& &- \frac{\lambda}{4\pi G} \int_{\partial M} d^4x \sqrt{-h} \,n_M
\epsilon^{MNPQR}D_N \xi K_{PL}D_Q K^L_R \,.
\end{eqnarray}

Now without lost of generality we can choose the gauge $N=1$ and
$N_A=0$ which define the so called Gaussian normal coordinates, and
the metric takes the form $ds^2 = dr^2 + \gamma_{ij} dx^i dx^j$. After
doing the decomposition in terms of surface induced and orthogonal
fields, all the terms depending on the extrinsic curvature cancel
thanks to the contributions from $S_{CSK}$! The gauge variation of the
action depends only on the intrinsic four dimensional curvature of the
boundary and is given by
\begin{equation}
 \delta S = \frac{1}{16 \pi G} \int_{\partial M} d^4 x \sqrt{-h}
\epsilon^{mnkl}\left( \frac{\kappa}{3} \hat{F}_{mn} \hat{F}_{kl} + \lambda \hat
R^{i}\,_{jmn} \hat R^{j}\,_{ikl}\right) \,.
\end{equation}
This has to be interpreted as the anomalous variation of the effective quantum
action of the dual field theory. As consequence of the discussion in the
subsection \ref{subsec:2} we can recognize  the form of the consistent anomaly
and use eq.~(\ref{eq:localanomaly})  to fix $\kappa$ for a single fermion
transforming under a $U(1)_L$ symmetry. Similarly we can fix $\lambda$ by
matching to the gravitational
anomaly of a single left-handed fermion eq.~(\ref{eq:gravanomaly}) and find
\begin{equation}
-\frac{\kappa}{48 \pi G} = \frac{1}{96 \pi^2}\quad ,\quad -\frac{\lambda}{16\pi
G} = \frac{1}{768 \pi^2}\, . 
\end{equation}

The bulk equations of motion are
\begin{eqnarray}\label{eqgrav}
 G_{MN} - \Lambda g_{MN} &=& \frac 1 2 F_{ML} F_N\,^L - \frac 1 8 F^2 g_{MN} + 2
\lambda \epsilon_{LPQR(M} \nabla_B\left( F^{PL} R^B\,_{N)}\,^{QR} \right) \,,
\label{eq:Gbulk}\\\label{eqgauge}
\nabla_NF^{NM} &=& - \epsilon^{MNPQR} \left( \kappa F_{NP} F_{QR} + \lambda 
R^A\,_{BNP} R^B\,_{AQR}\right) \,.  \label{eq:Abulk}
\end{eqnarray}

A remarkable fact is that the mixed Chern-Simons term does not
introduce new singularities into the on-shell action for any asymptotically
$AdS$ solution, i.e. no new counterterm is needed to renormalize the theory.
See~\cite{Landsteiner:2011iq} for a detailed discussion of the renormalization
of the model and Appendix~1 to see the counterterms. 
  
 \subsection{Applying Kubo Formulae and Linear Response}
 
In order to compute the conductivities under study using the Kubo formulae
eq.~(\ref{Kformulae}), we will use tools of linear response theory. To do so we
introduce metric and gauge fluctuations over a charged black hole
background and use the AdS/CFT dictionary to compute the retarded propagators
\cite{Son:2002sd,Herzog:2002pc}. Therefore we split the backgrounds and
fluctuations as,
\begin{eqnarray}
g_{MN} &=& g^{(0)}_{MN} + \epsilon \, h_{MN} \,,\\ 
A_{M} &=& A^{(0)}_{M} + \epsilon \, a_{M} \, .
\end{eqnarray}
After the insertion of these fluctuations and background fields in the action
and expanding up to second order in $\epsilon$ we can read the on-shell boundary
second order action which is needed to get the desired
propagators~\cite{Kaminski:2009dh},
\begin{equation}\label{eq:2ndor}
\delta S^{(2)}_{ren}=\int \frac{\mathrm d ^d k}{(2\pi)^d} \lbrace \Phi^I_{-k}
\mathcal A_{IJ} \Phi '^J_k + \Phi^I_{-k}  \mathcal B_{IJ} \Phi^J_k
\rbrace\Big{|}_{r\to\infty}\,,
\end{equation}
where prime means derivative with respect to the radial coordinate, $\Phi^I_{k}$
is a vector constructed with the Fourier transformed components of $a_M$ and
$h_{MN}$,
\begin{equation}
\Phi^I(r,x^{\mu})=\int \frac{\mathrm d^d k}{(2\pi)^d} \Phi^I_k (r) e^{-i \omega
t+i \vec{k}\vec{x}} \,,
\end{equation}
and $\mathcal{A}$ and $\mathcal{B}$ are two matrices extracted from the boundary
action and that we will show below.

For a coupled system the holographic computation of the correlators consists
in finding a
maximal set of linearly independent solutions that satisfy infalling
boundary conditions on the horizon and that source a single operator
at the AdS boundary
\cite{Son:2002sd,Herzog:2002pc,Kaminski:2009dh,Amado:2009ts}. To do so we can
construct a matrix of solutions
$F^I\,_J (k,r)$ such that each of its columns corresponds to one of
the independent solutions and normalize it to the unit matrix at the
boundary. Therefore, given a set of boundary values for the
perturbations, $\varphi^I_k$, the bulk solutions are 
\begin{equation}\label{eq:f}
\Phi^I_k (r) = F^I\,_J (k,r)\, \varphi^J_k\,.  
\end{equation}
Finally using this
decomposition we obtain the matrix of retarded Green's functions
\begin{equation}\label{eq:GR} 
G_{IJ}(k)= -2 \lim_{r\to\infty} \left(\mathcal A_{IM}
(F^M\,_J (k,r))' +\mathcal B_{IJ}\right)\, .  
\end{equation}

The system of equations (\ref{eqgrav})-(\ref{eqgauge}) admit the
following exact background AdS Reissner-Nordstr\"om black-brane
solution
 \begin{eqnarray}
 \mathrm d  s^2&=& \frac{r^2}{L^2}\left(-f(r) \mathrm d  t^2
+\mathrm d  \vec{x}^2\right)+\frac{L^2}{r^2 f(r)} \mathrm d r^2\,,\\ 
\label{eq:backA}
A^{(0)}&=&\phi(r)\mathrm d  t = \left(\nu - \frac{\mu \, r_{{\rm
H}}^2}{r^2}\right)\mathrm d  t\,,
\end{eqnarray}
where the horizon of the black hole is located at $r=r_{\rm H}$ and the
blackening factor of the metric is 
\begin{equation}f(r)=1-\frac{M L^2}{ r^4}+\frac{Q^2 L^2}{r^6}\,.  
\end{equation}

The parameters $M$ and $Q$ of the RN black hole are related to the
chemical potential $\mu$ and the horizon $r_H$ by~\footnote{The
  chemical potential is introduced as the energy needed to introduce
  an unit of charge from the boundary to behind the horizon
  $A(\infty)-A(r_{\mathrm H})$ which corresponds to the prescription
  (B) in table~\ref{tab:formalisms}. Observe that we have left the
  source value $A(\infty) =\nu$ as an arbitrary constant for reasons
  we will explain later. }
\begin{equation}
M=\frac{r_{\rm
    H}^4}{L^2}+\frac{Q^2}{r_{\rm H}^2}\quad,\quad Q=\frac{\mu\,
  r_{\rm H}^2}{\sqrt{3}}\,.  
  \end{equation}
The Hawking temperature is given in terms of these black hole parameters as 
\begin{equation}T=\frac{r_{\rm
    H}^2}{4\pi\, L^2} f^\prime(r_{\rm H}) = \frac{ \left(2\, r_{\rm H}^2\, M -
3\, Q^2 \right)}{2 \pi \,r_{\rm H}^5} \,.
\end{equation}
The pressure of the gauge theory is $P=\frac{M}{16\pi GL^3}$ and its energy
density is $\epsilon=3P$ due to the underlying conformal symmetry.

To study the effect of anomalies  we just  turned on the shear sector
(transverse momentum fluctuations) $a_\alpha$ and $h^\alpha_{\,t}$ and set
without loss of generality the momentum $k$ in the $y$-direction at zero
frequency, so $\alpha=x,z$. Since we are interested in a hydrodynamical regime
($k,\omega \ll T$), it is just necessary to find solutions up to first order in
momentum. So that  we expand the fields in terms of the dimensionless momentum
$p=k/4\pi T$ such as 
\begin{eqnarray}
h^\alpha_t(r) &=& h^{(0)\alpha}_t(r)+p \,h^{(1)\alpha}_t(r)
\,,\\\label{eq:redefB} B_\alpha(r) &=& B^{(0)}_\alpha(r)+p \,B^{(1)}_\alpha(r)\,
,  
\end{eqnarray}
 with the gauge field redefined as $B_\alpha=a_\alpha/\mu$.
For convenience we redefine new parameters and radial coordinate
\begin{equation}
\bar\lambda=\frac{4\mu \lambda L}{r_H^2}\quad;\qquad \bar \kappa=\frac{4\mu
\kappa L^3}{r_H^2}\quad;\qquad a=\frac{\mu^2L^2}{3r_{\rm H}^2}\quad;\qquad
u=\frac{r_H^2}{r^2}\,.
\end{equation}
In this new radial coordinate the horizon sits at $u=1$ and the $AdS$
boundary at $u=0$. At zero frequency the system of differential
equations consist on four second order equations.\footnote{The
  complete system of equations depending on frequency and momentum is
  showed in Appendix~2. The system consists of six dynamical equations
  and two constraints.} The relevant physical boundary conditions on
fields are: $h^\alpha_t(0)=\tilde H^\alpha$, $B_\alpha(0)=\tilde
B_\alpha$; where the `tilde' parameters are the sources of the
boundary operators.  The second condition compatible with the ingoing
one at the horizon is regularity for the gauge field and vanishing for
the metric fluctuation \cite{Amado:2011zx}.

After solving the system perturbatively (see \cite{Landsteiner:2011iq} for
solutions), we can go back to the formula (\ref{eq:GR}) and compute the
corresponding holographic Green's functions. 
If we consider the vector of fields to be 
\begin{equation}
\Phi_k^{\top} (u) = \Big{(} B_x(u) ,\, h^x_{\,t}(u)  ,\, B_z(u) ,\, h^z_{\,t}(u)
\Big{)} \,,
\end{equation}
the $\mathcal A$ and $\mathcal B$ matrices for that setup take the following
form

\begin{equation}
\mathcal A =\frac{r_{\rm H}^4}{16\pi G L^5} \,{\rm Diag}\left( -3a f,\,
\frac{1}{u} ,\, -3a f,\, \frac{1}{u} \right) \,,
\end{equation}

\begin{eqnarray}
\hspace{-0.8cm}\mathcal B _{AdS+\partial}=
\frac{r_{\rm H}^4}{16\pi G L^5}
\left(
\begin{array}{cccc}
0 & -3a  & \frac{4 \kappa i k \mu^2 \phi L^5}{3r_{\rm H}^4} & 0  \\
0 & -\frac{3}{u^2} &  0 & 0  \\
\frac{-4 \kappa i k \mu^2 \phi L^5}{3r_{\rm H}^4} & 0 & 0 & -3a  \\
0 & 0 & 0 & -\frac{3}{u^2}  \\
\end{array}
\right)\,,
\end{eqnarray}

\begin{eqnarray}
\mathcal B _{CT}=
\frac{r_{\rm H}^4}{16\pi G L^5}
\left(
\begin{array}{cccc}
\vspace{0.15cm}
0&  0 & 0 & 0 \\
\vspace{0.15cm}
0 & \frac{3}{u^2 \sqrt{f\,}} & 0 & 0  \\
\vspace{0.15cm}
0 & 0  & 0 & 0\\
\vspace{0.15cm}
0 & 0  & 0 & \frac{3}{u^2 \sqrt{f\,}}  \\
\end{array}
\right)\,,
\hspace{-0.8cm}\,
\end{eqnarray}
where $\mathcal B =\mathcal B _{AdS+\partial}+\mathcal B
_{CT}$.~\footnote{$\mathcal{B}_{CT}$ is coming from the counterterms of the
theory.} Notice that there is no contribution to the matrices coming from the
Chern-Simons gravity part, because the corresponding contributions vanish at the
boundary. These matrices and the perturbative solutions are the ingredients to
compute the matrix of propagators. Undoing the vector field redefinition
introduced in~(\ref{eq:redefB}) the non-vanishing retarded correlation functions
at zero frequency are then
\begin{eqnarray}
\label{eq:giti}G_{x,tx} &=& G_{z,tz} =\frac{\sqrt{3}\, Q}{4 \pi\, G\, L^3  }\,,
\\
\label{eq:gxz}G_{x,z} &=& - G_{z,x} =\frac{i\, \sqrt{3}\, k\, Q\,  \kappa}{2
\pi\, G \, r_{\rm{H}}^2}+\frac{i\, k\, \nu \, \kappa}{6\pi\, G  }\,,
\end{eqnarray}
\begin{eqnarray}
G_{x,tz} &=&G_{tx,z} = -G_{z,tx}=-G_{tz,x}=\frac{3\, i\, k\, Q^2\,  \kappa }{4
\pi\,G\,  r_{\rm{H}}^4}+\frac{2ik\lambda \pi T^2}{G} \,, \\
G_{tx,tx} &=& G_{tz,tz}=\frac{M}{16\pi\, G\, L^3 }\,,\\
\label{eq:gtiti}G_{tx,tz} &=& -G_{tz,tx}=+\frac{i\, \sqrt{3}\, k\, Q^3\,
\kappa}{2\pi\, G\,r_{\rm{H}}^6} + \frac{4\pi i\sqrt{3}  k Q T^2 \lambda}{G
\,r_{\rm H}^2}\,.
\end{eqnarray}

We can do an important remark observing eq.~(\ref{eq:gxz}). Remember that we
left the boundary value of the background gauge field eq.~(\ref{eq:backA})
arbitrary as a constant $\nu$. But as the $U(1)$ symmetry is anomalous in the
Field Theory side, physical quantities have to be sensitive to the source
$A_0$,~\footnote{In principle $A_0$ could be gauged away for the symmetric case
and in consequence observables should not depend on its value. For example look
at \cite{Gynther:2010ed} to see how in presence of a $U(1)_V\times U(1)_A$
symmetry  with only the $U(1)_V$ conserved, propagators do not depend on the
specific value of the zero component of the vector gauge source $V_0$.} indeed
as we can check they are. In particular if we choose the  value $\nu=\mu$ which
corresponds to formalism (A) in table~\ref{tab:formalisms}, we need to include
the counterterm eq.~(\ref{eq:Stheta}) in order to get the same propagator as at
weak coupling. In fact in \cite{Rebhan:2009vc,Gynther:2010ed} it has been shown
that in the case of a propagator between  two vector currents, choosing this
specific value for $\nu$ the propagator would be zero, giving us in consequence
a zero value for the chiral magnetic conductivity. Hence to be consistent with
the scheme  we are working at, let us just consider $\nu$ as a source in the
field theory. Therefore the real propagator is the one with $\nu=0$ because as
is well known we have to set all sources to zero after taking the second
functional derivative of the efective action. Finally using the Kubo formulae
(\ref{Kformulae}) we recover the vortical and axial-magnetic conductivities  
\begin{eqnarray}
\label{eq:sigb}\sigma_B &=& -\frac{\sqrt{3}\,  Q \, \kappa}{2 \pi
  \,G\, r_{\rm{H}}^2} =  \frac{ \mu}{4 \pi^2}\,,\\
\label{eq:sigv}\sigma_V &=&\sigma_B^\epsilon=-\frac{3\, Q^2\,  \kappa }{4
\pi\,G\, \bar  r_{\rm{H}}^4}-\frac{2\lambda \pi T^2}{G}= \frac{\mu^2}{8\pi^2}
+\frac{T^2}{24} \,, \\
\label{eq:sigve}\sigma_V^\epsilon &=&-\frac{ \sqrt{3}\, Q^3\,
  \kappa}{2\pi\, G\, r_{\rm{H}}^6} - \frac{4\pi \sqrt{3}   Q T^2
  \lambda}{G \,r_{\rm H}^2} =\frac{\mu^3}{12 \pi^2}+\frac{\mu
  T^2}{12}\,. 
\end{eqnarray}
All these expressions coincide with the results in
section~\ref{sec:3}, (\ref{eq:cmclab}), (\ref{eq:cvclab}) and (\ref{eq:cvcelab})
if we
specialize to $d_{abc}=1$ and $b_a=1$. They are in perfect agreement with the
literature
\cite{Erdmenger:2008rm,Banerjee:2008th,Son:2009tf,Amado:2011zx} except
for the contribution coming from the gravitational anomaly which is
manifest by the presence of the extra~$\lambda T^2$. All the numerical
coefficients coincide precisely with the ones obtained at weak
coupling; this we take as a strong hint that the anomalous
conductivities are indeed completely determined by the anomalies and
are not renormalized beyond one loop. Evidence for non-renormalization
comes also from~\cite{Landsteiner:2012dm} where a holographic renormalization
group running of the conductivities showed the same result at any
value of the holographic cut-off.  We also point out that the $T^3$ term that
appears as
undetermined integration constant in the hydrodynamic considerations
in \cite{Neiman:2011mj} should make its appearance in
$\sigma_V^\epsilon$. We do not find any such term which is consistent
with the argument that this term is absent due to CPT invariance.

It is also interesting to write down the magnetic and vortical
conductivites using eqs.~(\ref{eq:cmc}) and (\ref{eq:cvc}) as they appear  in
the Landau frame to compare with the Son and Surowka form  \cite{Son:2009tf} 
\begin{eqnarray}
\label{eq:xib}\xi_B &=& \textstyle-\frac{\sqrt{3} Q ( ML^2+3 r_{\rm
H}^4)\kappa}{8\pi G M L^2 r_{\rm H}^2}+\frac{\sqrt{3}Q\lambda\pi T^2}{GM} =
\frac{1}{4\pi^2}\, \left( \mu -  \frac{1}{2}\frac{n(\mu^2+\frac{\pi^2
T^2}{3})}{\epsilon+P}\right)\,, \\
\label{eq:xiv} \xi_V &=&\textstyle-\frac{3 Q^2  \kappa }{4\pi G  ML^2
}-\frac{2\pi\lambda T^2(r_{\rm H}^6-2L^2Q^2)}{ G M L^2 r_{\rm H}^2} =  
\frac{\mu^2}{8\pi^2}\, \left( 1 - \frac{2}{3}
  \frac{n \mu}{\epsilon+P}\right)+\frac{T^2}{24}\left(1-\frac{2
n\mu}{\epsilon+P}\right)\,.
\end{eqnarray}
These expressions agree with the literature except for the $\lambda T^2$ term. A
last comment can be done, the shear viscosity entropy ratio is not modified by
the presence of the gravitational anomaly. We know that
$\eta\propto\lim_{\omega\to 0}\frac{1}{\omega}<T^{xy}T^{xy}>_{k=0}$, so we
should solve the system at $k=0$ for the fluctuations $h^i_y$  but the 
anomalous coefficients always appear with a momentum $k$ (see Appendix~2),
therefore if we switch off the momentum, the system looks precisely as the
theory without anomalies. In \cite{Bonora:2011mf} it has been shown that the
black hole entropy doesn't depend on the extra Chern-Simons term.~\footnote{For
a four dimensional
holographic model with gravitational Chern-Simons term and a scalar field this
has also been shown in \cite{Delsate:2011qp}.}

\section{Conclusion and Outlook}
\label{sec:6}

In the presence of external sources for the energy momentum tensor and
the currents, the anomaly is responsible for a non conservation of the
latter. This is conveniently expressed
through~\cite{AlvarezGaume:1983ig}
\begin{equation}
 D_\mu J_a^\mu= \epsilon^{\mu\nu\rho\lambda}\left( \frac{d_{abc}}{32\pi^2} 
F^b_{\mu\nu} F^c_{\rho\lambda} + \frac{b_a}{768\pi^2} 
R^\alpha\,_{\beta\mu\nu}
R^\beta\,_{\alpha\rho\lambda}\right) \,,
\end{equation}
where the axial and mixed gauge-gravitational anomaly coefficients, $d_{abc}$
and $b_a$, are given by~(\ref{eq:chiralcoeff}) and (\ref{eq:gravcoeff})
respectively.

We have discussed in Sec.~\ref{sec:2} the constitutive relations and
derived the Kubo formulae that allow the calculation of transport
coefficients at first order in the hydrodynamic expansion. We
explained also subtleties in the definition of the chemical potential
in the presence of anomalies. The explicit evaluation of these Kubo
formulae in quantum field theory has been performed in
Sec.~\ref{sec:3} for the chiral magnetic, chiral vortical and energy
flux conductivities of a relativistic fluid at weak coupling, and we
found contributions proportional to the anomaly coefficients~$d_{abc}$
and $b_a$. Non-zero values of these coefficients are a necessary and
sufficient condition for the presence of anomalies
\cite{AlvarezGaume:1983ig}.  Therefore the non-vanishing values of the
transport coefficients have to be attributed to the presence of chiral
and gravitational anomalies.

In order to perform the analysis at strong coupling via AdS/CFT methods, we
have defined in Sec.~\ref{sec:4} a holographic model implementing
both type of anomalies via gauge
and mixed gauge-gravitational Chern-Simons terms. We have computed the
anomalous magnetic and vortical conductivities from a charged black
hole background and have found a non-vanishing vortical conductivity
proportional to $\sim T^2$. These terms are characteristic for the
contribution of the gravitational anomaly and they even appear in an
uncharged fluid. The~$T^2$ behavior had appeared already previously in
neutrino physics~\cite{Vilenkin:1979ui}. In~\cite{Neiman:2010zi}
similar terms in the vortical conductivities have been argued for, but
just in terms of undetermined integration constants without any
relation to the gravitational anomaly. Very recently  a generalization of the
results ~(\ref{eq:sigb})-(\ref{eq:sigve}) to any even space-time dimension as
a polynomial in~$\mu$ and $T$~\cite{Loganayagam:2012pz} has been
proposed. Finally, the
consequences of this anomaly in hydrodynamics have been studied using
a group theoretic approach, which seems to suggest that their effects
could be present even at $T=0$~\cite{Nair:2011mk}.  The
numerical values of the anomalous conductivities at strong coupling
are in perfect agreement with weak coupling calculations, and this
suggests the existence of a non-renormalization theorem including the
contributions from the gravitational anomaly.

There are important phenomenological consequencies of the present
study to heavy ion physics. In~\cite{KerenZur:2010zw} enhanced
production of high spin hadrons (especially $\Omega^-$ baryons)
perpendicular to the reaction plane in heavy ion collisions has been
proposed as an observational signature for the chiral separation
effect. Three sources of chiral separation have been identified: the
anomaly in vacuum, the magnetic and the vortical conductivities of the
axial current $J_A^\mu$. Of these the contribution of the vortical
effect was judged to be subleading by a relative factor of
$10^{-4}$. The~$T^2$ term in (\ref{eq:sigv}) leads however to a
significant enhancement. If we take $\mu$ to be the baryon chemical
potential $\mu\approx 10$ MeV, neglect $\mu_A$ as in
\cite{KerenZur:2010zw} and take a typical RHIC temperature of $T=350$
MeV, we see that the temperature enhances the axial chiral vortical
conductivity by a factor of the order of $10^4$. We expect the
enhancement at the LHC to be even higher due to the higher
temperature.

In this review we have presented the computation of the transport coefficients,
and in particular their gravitational anomaly contributions, via Kubo formulae.
It would be interesting to calculate directly the constitutive relations of the
hydrodynamics of anomalous currents via the fluid/gravity correspondence within
the holographic model of Sec.~\ref{sec:4},
~\cite{Erdmenger:2008rm,Banerjee:2008th,Bhattacharyya:2008jc}. This approach
will allow us to compute transport coefficients at higher
orders~\cite{Kharzeev:2011ds,Bhattacharyya:2012ex}. This study is currently in
progress~\cite{workinprogress}.

\begin{acknowledgement}
This work has been supported by Plan Nacional de Altas Energ\'{\i}as
FPA2009-07908, FPA2008-01430 and FPA2011-25948, CPAN (CSD2007-00042),
Comunidad de Madrid HEP-HACOS S2009/ESP-1473. E.~Meg\'{\i}as would
like to thank the Institute for Nuclear Theory at the University of
Washington, USA, and the Institut f\"ur Theoretische Physik at the
Technische Universit\"at Wien, Austria, for their hospitality and
partial support during the completion of this work. The research of
E.M. is supported by the Juan de la Cierva Program of the Spanish
MICINN. F.P. has been supported by fellowship FPI Comunidad de Madrid.
\end{acknowledgement}
%

\section*{Appendix 1: Boundary Counterterms }
\addcontentsline{toc}{section}{Appendix}
\label{app:boundary_counterterms}

The result one gets for the counterterm coming from the regularization of the
boundary action of the holographic model in section~\ref{sec:4} is
\begin{equation}
\!S_{ct} = - \frac{3}{8\pi G} \int_{\partial M} d^4x \sqrt{-h} \bigg[
1 + \frac{1}{2}P  - \frac{1}{12} \left( P^i_j P^j_i - P^2 -  \frac{1}{4}
\hat{F}_{ij} \hat{F}\,^{ij} \right)\log e^{-2\rho} \bigg] \,, \label{eq:Sct}
\end{equation}
where hat on the fields means the induced field on the cut-off surface and
\begin{equation}
P = \frac{1}{6}\hat{R} \,, \qquad  P^i_j = \frac{1}{2} \left[ \hat{R}^i_j - P
\delta^i_j \right] \,. \label{eq:PPij} 
\end{equation}
As a remarkable fact there is no contribution in the counterterm
coming from the gauge-gravitational Chern-Simons term. This has also
been derived in \cite{Clark:2010fs} in a similar model that does
however not contain $S_{CSK}$.

\section*{Appendix 2: Equations of motion for the shear sector}
\addcontentsline{toc}{section}{Appendix}
\label{eq_shear}
These are the complete linearized set of six dynamical equations of motion,
\begin{eqnarray}
\label{eq_As}\nonumber 0&=& B_\alpha'' ( u )+\frac { f' ( u ) }{f(u)}B_\alpha'(
u )+ \frac {b^2  }{uf( u )^2 }\left(\omega^2-  f( u) 
{k}^{2} \right)B_\alpha( u) -  \frac{ h^{\alpha'}_t ( u
 )}{f(u)} \\
 &&+ik\epsilon_{\alpha\beta}\left(\ \frac{3}{u f(u)} \bar\lambda \left(
\frac{2}{3a} (f(u)-1)+ u^3 \right) h_t^{\beta'}(u)+\bar \kappa
\frac{B_\beta(u)}{f(u)} \right) \,,\\
\label{eq_Hts}\nonumber 0&=&  h^{\alpha''}_t(u) - \frac{h^{\alpha'}_t(u)}{u}
-\frac {b^2}{
uf( u )}\left(k^2 h^\alpha_t(u)+ h^\alpha_y \left( u \right) \omega k \right)
- 3a u B'_\alpha(u) \\
\nonumber &&i\bar\lambda k \epsilon_{\alpha\beta}\left[\left(24a
u^3-6(1-f(u))\right)\frac {B_\beta(u)}{u}+(9a u^3-6(1-f(u)))B'_\beta(u)\right.\\
 &&\left.+2 u (uh^{\beta'}_t ( u ) )'  -\frac {  2u  b^2}{  f ( u) }\left(
h_y^\beta( u ) \omega k+h_t^\beta( u ) k^2\right)\right] \,,
\\
\label{eq_Hys}\nonumber 0&=& h^{\alpha''}_y(u)+
\frac{\left(f/u\right)'}{f/u}h^{\alpha'}_y(u)+\frac{b^2}{u
f(u)^2}\left(\omega^2h_y^\alpha(u) + \omega k
h_t^\alpha(u)\right)+2uik\bar{\lambda}\epsilon_{\alpha\beta}\left[u 
h^{\beta''}_y(u)\right.\\
 && \left.+\left(9f(u)-6+3a u^3\right)\frac{h^{\beta'}_y(u)  }{f(u)}+\frac{ b^2
}{f(u)^2}\left(\omega k h_t^\beta(u)+w^2 h_y^\beta(u)\right)\right] \,,
\end{eqnarray}
and two constraints for the fluctuations at $\omega ,k\neq 0$ 
\begin{eqnarray}
\label{constraints}\nonumber 0&=&  \omega \left(h^{\alpha'}_t(u)-3a u
B_\alpha(u)\right)+f(u)kh^{\alpha'}_y(u)+i
k\bar\lambda\epsilon_{\alpha\beta}\left[2u^2 \left(\omega h^{\beta'}_t+ f( u ) 
k h^{\beta'}_y (u)\right)\right.\\
 && \left. +\left(9a u^3-6(1-f(u))\right)B_\beta(u)\right] \,.
\end{eqnarray}

\end{document}